# Venus as an Exoplanet: I. An Initial Exploration of the 3-D Energy Balance for a $CO_2$ Exoplanetary Atmosphere Around an M-Dwarf Star

C. D. Parkinson[1, 2, *], S. W. Bougher[3], F. P. Mills[1,4,5], A. Brecht[6], J. Li[4],

Y. L. Yung[2,7], R. Hu[7], and G. Gronoff[8]

[1]*Space Science Institute*

*4765 Walnut St, Suite B*

*Boulder, CO, 80301, USA*

[2]*Division of Geological and Planetary Sciences*

*California Institute of Technology, CA, USA*

[3]*Climate and Space Sciences and Engineering Department*

*University of Michigan*

*2455 Hayward Street, Ann Arbor, MI, 48109, USA*

[4]*Fenner School of Environment & Society*

*Australian National University, Canberra, Australia*

[5]*McDonald Observatory*

*University of Texas Austin*

*Austin, TX, 78712 USA*

[6]*NASA Ames Research Center, Moffett Field, CA, USA*

[7]*Jet Propulsion Laboratory, CA, USA*

[8]*LARC-E303, Science Systems & Applications, Inc*

[*] *To whom correspondence should be addressed*

*email: cparkinson@spacescience.org*

*Key Points:*

1. A 3-D GCM simulates middle and upper atmospheric effects of energy balance and dynamics for a Venus-type exoplanet orbiting GJ 436.
2. Increasing stellar EUV/UV fluxes for decreasing planet-star distances produce increasing zonal wind, thermospheric temperature, and conductive cooling.
3. Energy balance sensitivity study shows that (a) a ±30% change in the NIR heating profile only incrementally changes the thermal structure above $10^{-5}$ mbar, and virtually no change below $10^{-5}$ mbar, and (b) doubling and quadrupling the NIR heating profile produces significant changes in dynamics and temperature throughout the range of atmospheric pressures considered.
4. $CO_2$ 15-μm cooling is a strong thermostat regulating dayside temperatures due to cooling enhancement via collisions of O and $CO_2$.

**Plain Language Summary**

Understanding the state and composition of an exoplanetary atmosphere depends upon several parameters such as heating, cooling, mixing and reactions between constituent chemical species. Only a few types of atmospheric species can be detected remotely via spectroscopy and only if their abundance is large enough to be detectable. In this initial study, we model the atmosphere of a Venus-like planet going around the M-type star GJ 436 to determine the global neutral temperature structure, winds, and energy balance as the radial distance of the planet from the star decreases.

**Abstract**

The chemical evolution of an exoplanetary Venus-like atmosphere is dependent upon the ultraviolet to near ultraviolet (FUV-NUV) radiation ratio from the parent star, the balance between $CO_2$ photolysis and recombination via reactions that depend on the water abundance, and various catalytic chemical cycles. In this study, we use a three-dimensional (3-D) model to simulate conditions for a Venus-like exoplanet orbiting the M-dwarf type star GJ436 by varying the star/planet distance and considering the resultant effects on heating/cooling and dynamics. The simulation includes the middle and upper atmosphere (<40 mbar). Overall, these model comparisons reveal that the impact of extreme ultraviolet to ultraviolet (EUV-UV) heating on the energy balance shows both radiative and dynamical processes are responsible for driving significant variations in zonal winds and global temperature profiles at $< 10^{-5}$ mbar. More specifically, $CO_2$ 15-μm cooling balances EUV/UV and Near InfraRed (NIR) heating at altitudes below $10^{-7}$ mbar pressure with a strong maximum balance for pressures at $\sim 10^{-5}$ mbar, thus explaining the invariance of the temperature distribution at altitudes below $10^{-5}$ mbar pressure for all cases. Our model comparisons also show that moderate changes in NIR heating result in relatively small changes in neutral temperature in the upper atmosphere, and virtually no change in the middle atmosphere. However, with larger changes in the NIR heating profile, much greater changes in neutral temperature occur in the entire upper and middle atmosphere studied.

## *1. Introduction*

Atmospheric motions are an inescapable response to a planet's various forces such as rotation, gravity, and differential stellar heating of the atmosphere. Dynamics constitutes the relationships between the heating, forces, and winds that they drive. Dynamics can also affect a planet's climate evolution and habitability. Winds which are variable in time transport heat, which in turn affects atmospheric chemistry. Movement of air carries condensable species responsible for clouds, such as water on Earth and Mars, ammonia on Jupiter, methane on Titan, or sulfuric acid on Venus. Dynamics also strongly affects the distribution of non-condensable trace gases. Generally, winds have a preferred direction that either persists seasonally or annually. Such average winds make up the general circulation of a planetary atmosphere. The role of atmospheric dynamics in the long-term evolution of atmospheres and planetary habitability remains relatively unexplored. Interpretation of forthcoming terrestrial-like exoplanet observations and assessment of their implications for comparative planetary evolution requires understanding the chemical and physical processes operating in exoplanetary atmospheres.

Venus observations and models are two important sources for relevant information when studying $CO_2$ dominated exoplanetary atmospheres. The theoretical understanding of these worlds is critical for the operation of observatories such as the James Webb Space Telescope and the future large (> 30m) ground-based observatories that will have an objective to study the habitability of planets. First, because Venus-like planets may be targets of interest in themselves (Kane et al., 2014, 2018, 2019, 2021; Way et al., 2016, 2020), but overall because it is important to understand why Venus is so different from the Earth in terms of evolution. Hence, the objective of several JWST observations is to observe Venus-like atmospheres and understanding the stability of these atmospheres in the light of their host-stars is crucial. The thermal stability, partially driven by the wind patterns, is an also essential point to address. The observation of Venus-like exoplanets will also help better understand the history of the Earth, since the same processes that led to Venus' atmosphere may have helped for the early habitability of the Earth (Turbet et al., 2021).

Several types of models have been used to study terrestrial-like exoplanets. 1-D photochemical (*e.g.*, Harman et al., 2018) and radiative-convective (*e.g.*, Robinson and Crisp, 2018; Lincowski et al., 2018; Meadows et al., 2018; Mickol et al., 2015; Robinson and Caitlin, 2012) models have been employed to simulate the global-average atmospheric composition and evolution of Venus-like exoplanets, including the upper atmospheric conditions and composition that will dominate observed spectra. 3-D GCMs (*e.g.*, Way & Del Genio, 2020) have been employed to simulate their bulk atmospheres.

Previous work using 1-D photochemical models has been published looking at the impact of the planet-star distance for Venus-like and/or Earth-like atmospheric compositions (Arney et al. 2017; Arney, 2019; Lincowski et al. 2018; Hu et al. 2020). 1-D photochemical models have been used frequently owing to the large potential parameter space (*e.g.*, Gao et al., 2015; Segura et al., 2005, 2007; Domagal-Goldman et al., 2014). While ideal for this purpose, these models have important inherent limitations. In particular, a global-/diurnal-average simulation is not the same as what one would get from averaging over the global state of an atmosphere (*e.g.*, Mills et al, 2021), it's not possible to simulate the terminator region that a transit observation would sense, and dynamical influences on species distributions are ignored. 1-D radiative convective simulations can parameterize the effects of condensable species and clouds but suffer the same spatial limitations as 1-D photochemical models. 3-D General Circulation Models (GCMs) have the ability to interpret and connect observations due to the multi-dimensions being represented.

Exoplanet climate simulations have been done previously with the Laboratoire de Meterorologie Dynamique-generic (LMD-G) Global Climate Model (GCM; Wordsworth et al., 2011) (*e.g.*, Pidhorodetska et al., 2020; Fauchez et al., 2019; Turbet et al., 2018, Arney et al., 2016, 2017). Details on the LMD-G GCM can be found in Turbet et al. (2018) and Fauchez et al. (2019).

Like most GCMs used in exoplanet atmospheric research, the LMD-G GCM does not include photochemistry (Pidhorodetska et al., 2020). To simulate atmospheric composition a 1-D photochemistry code is utilized to produce vertical chemical profiles separate from the LMD-G GCM. See details in Arney et al. (2016, 2017); Lincowski et al. (2018); Meadows et al. (2018), and review in Parkinson et al. 2021.

The 3-D Venus Thermospheric General Circulation Model (VTGCM) was designed for the study of Venus' middle and upper atmosphere (*e.g.,* Parkinson et al., 2015a, b, 2021; Bougher et al., 2015; Brecht et al., 2011, 2021) and so is ideally suited for exploring the portion of an exoplanetary atmosphere typically probed by transmission spectroscopy. In this initial study, we use the well-tested and validated VTGCM to simulate middle and upper atmosphere conditions on a Venus-like planet at varying distances from the M-dwarf GJ 436 and compare the results with those previously published for Venus. We believe these are the first 3-D simulations with self-consistent dynamics, energetics, and photochemistry focused on the observable middle and upper atmosphere of terrestrial exoplanets having thick $CO_2$-dominated atmospheres (i.e., Venus-like exoplanets). Venus is a good proxy for an exoplanet study around M-dwarf stars since such exoplanets are likely tidally locked in their orbits and Venus has a very slow rotational period which approximates this well.

Several processes have been previously identified as important determinants of composition and chemistry in the observable upper atmosphere of an exoplanet: the ratio of FUV and NUV stellar radiation (e.g., Tian et al., 2014), total stellar FUV flux (e.g., Domagal-Goldman et al., 2014), and alternative catalytic chemical cycles (*e.g.*, Gao et al., 2015; Grenfell et al., 2013). We have incorporated these processes using the VTGCM to perform investigations using coupled atmospheric energetics (heating/cooling), dynamics, and photochemistry (*e.g.*, Parkinson et al., 2017).

### *1.1 The Influence of Stellar Characteristics*

Multiple stellar characteristics can profoundly influence a planetary atmosphere. For instance, an effect of flares associated with the creation of Stellar Energetic Particle (SEP) events can last for days and can affect the chemistry of the mesosphere and thermosphere (Airapetian et al., 2016, 2019; Hayworth et al. 2022). Flare events have not been studied for a dense $CO_2$-dominated atmosphere, but a study of their impact on a Mars-like atmosphere (Pawlowski and Ridley, 2008, 2009a, 2009b; Bougher et al., 2014) and an $O_2$-rich atmosphere found they are likely to only have short-term effects (Segura et al. 2010). We assume that the M-dwarf of this study is quiet enough to neglect the effects of these particles and solar flare fluxes. Consequently, they have not been included in this study, which focuses on the long-term steady-state.

The FUV to NUV ratio and absolute intensity of FUV radiation may also have strong effects on an atmosphere, but previous simulations have yielded conflicting results. Tian et al. (2014) suggested that a high FUV to NUV ratio was responsible for the buildup of $O_2$ because of decreased photolysis of $O_3$, $H_2O_2$, and $HO_2$ in their M-star case, but Harman et al. (2015) increased the FUV flux, while holding the NUV flux constant, and found that higher FUV fluxes increased $O_2$ concentrations only slightly. Both note $HO_2$ and $H_2O_2$ can be considered reservoirs for OH with the ultimate source being photolysis of $H_2O$. Harman et al. (2015) showed water vapor photolysis slows by a factor of 400 in their GJ 876 case, which has smaller FUV flux compared to their solar case, while the size of the $HO_2$ and $H_2O_2$ reservoirs are 2–3 orders of magnitude larger. However, neither appears to have analyzed the partitioning among $HO_x$ species to quantify dependences on FUV and NUV fluxes. For example, it is not clear how important the $O(^1D) + H_2O \rightarrow 2OH$ reaction is in the Tian et al. (2014) or Harman et al. (2015) calculations. This reaction is particularly important in the Earth's atmosphere with $O(^1D)$ produced by photolysis of $O_3$.

Similar unresolved differences exist regarding the impact of total FUV flux on $O_2$ and $O_3$ abundances. Domagal-Goldman et al. (2014) attributed the buildup of $O_3$ in their F-star case to the larger FUV flux, but Harman et al. (2015) found substantially more $O_2$ in their simulation of a planet around a K-star than for a planet subject to the higher FUV flux orbiting an F-star, the opposite dependence of that from Domagal-Goldman et al. (2014). Harman et al. (2015) had nearly the same FUV/NUV ratio for both the K- and F-star in their pair of simulations. A few works have been published after 2015 and seem to have largely resolved those issues. Particularly, Harman et al. (2018) and Hu et al. (2020). It is to be noted that the uncertainties in the $O_2$ buildup not only reside in the chemistry but also in the boundary conditions: the interaction with a potential magma-ocean, NO sinks (see Krissansen-Totton, et al., 2021 and references contained therein) have the potential to absorb the created oxygen, and conversely, atmospheric escape could lead to a large loss of $O_2$ (Gronoff et al. 2020 and references contained therein).

In the case of the current study, we assume that the host M-dwarf is quiet enough to not photodissociate all the $CO_2$ in the upper atmosphere through flares and particle precipitation; however, the stellar EUV-XUV flux is still considered. We will verify that, under these assumptions, enough $CO_2$ is present in the thermosphere to cool it down for negligible (*i.e.,* Venus-like) atmospheric escape.

GJ 436 is an M2.5V star whose characteristics have been observed extensively, partly due to the discovery of GJ 436b, a Neptune-size planet orbiting it (*e.g.*, Butler et al., 2004; von Braun et al., 2012). Its spectrum has been measured at UV (France et al., 2013) and Near InfraRed (NIR) (Terrien et al., 2015) wavelengths and its FUV/NUV ratio is high compared to that of the Sun. The GJ 436 spectrum has been used in several previous exoplanet simulations (*e.g.*, Segura et al., 2003; Gao et al., 2015, Harman et al., 2018). The actual exoplanet GJ 436b is thought to be in a state of hydrodynamic escape (Ehrenreich et al., 2015) since a large cloud of $H_2$ has been observed in its vicinity. However, this is a Neptune-like planet, yet in the case of a Venus-like atmosphere, the depletion of H means that we can have a more stable atmosphere with higher-mass constituents. As we will see in the following, the cooling by $CO_2$ prevents the atmosphere to be in a hydrodynamic-escape state. In 2012, although yet to be confirmed, NASA announced that astronomers had observed a second planet, GJ 436c (Sydney Morning Herald. Reuters. July 2012; Demory et al., 2009). It was measured to have a radius of around two thirds that of Earth (and Venus) and, assuming an Earth-like density of 5.5 g/cm$^3$, was estimated to have a mass of 0.3 times that of Earth and a surface gravity of around two thirds that of Earth. It orbits at 0.0185 AU from the star, every 1.3659 days.

*1.2 Heating, Cooling, and Dynamics*

The ability of photochemical processes to influence broad regions of an atmosphere depends on transport, which is controlled by differential heating and cooling in the atmosphere. On present day Venus (and ancient Mars), the dayside heat budget is (was) dominated by a balance of EUV heating and $CO_2$ 15-μm cooling above the near IR heating layer (*e.g.,* Bougher et al. 1999, 2002; Valeille et al. 2009; 2010; Parkinson et al., 2021). For present day Mars and Earth, $CO_2$ 15-μm cooling plays a different, but still important role in the heat budget (*e.g.*, Bougher et al., 2002). Photolysis of $CO_2$ and $O_2$ generates a large quantity of atomic O, and subsequent O-$CO_2$ collisions pump up the first $CO_2$ vibrational level, thereby enhancing $CO_2$ cooling where the timescale for collisional de-excitation exceeds the timescale for the emission of a 15-μm photon (*e.g.*, Bougher et al., 1999, 2017c; Parkinson et al., 2021). Spectra of GJ 436 (*cf.*, Figure 1) from the Hubble Space Telescope (HST) Measurements of the Ultraviolet Spectral Characteristics of Low-mass Exoplanetary Systems (MUSCLES) Treasury Survey indicate reduced stellar fluxes (with respect to our Sun) in the spectral region where atmospheric $CO_2$, $H_2O_2$, $O_2$, and $O_3$ photolysis occurs (Tian et al., 2014; Gao et al., 2015). $H_2O$, $O_2$, and $CO_2$ photolysis is driven by FUV photons whereas $H_2O_2$, $HO_2$, and $O_3$ photolysis is driven by NUV photons. The relative flux distributions as a function of spectral type described by Pickles (1998) show greatly reduced stellar fluxes for both K and M stars and increased stellar flux for F type stars (as compared to G type stars like the Sun). This should mean less atomic O production, leading to reduced $CO_2$ cooling and warmer upper atmosphere temperatures (above any near IR heating layer) for Venus-like planets around K and M class stars, despite their smaller fluxes. We discuss an exo-Venus preliminary modeling simulation result below in section 3, where we examine the impact of EUV-UV and NIR heating on the energy balance and dynamics for a $CO_2$ dominated exoplanetary atmosphere.

## 2. *Numerical Modeling Tool: VTGCM*

The VTGCM is a 3-D finite difference hydrodynamic model of Venus' upper atmosphere developed from NCAR's terrestrial Thermospheric Ionospheric GCM (TIGCM) (e.g., Bougher et al., 1988). The VTGCM solves the time-dependent primitive equations for the neutral upper atmosphere. Additionally, the prognostic equations (thermodynamic, eastward and northward momentum, composition) are typically solved for steady-state solutions for the temperature, zonal and meridional velocity, and the mass mixing ratios of specified major and minor species (*e.g.,* Brecht et al., 2011, 2021; Bougher et al., 2015). This model has been documented in detail as revisions and improvements have been made over more than three decades including modern parameterizations for $CO_2$ 15-μm cooling, near IR heating, wave drag, and eddy diffusion enabling the VTGCM to reproduce many *Pioneer Venus* and *Venus Express* (VEx) observations, including seasonal effects and nightglow intensities (*e.g.*, Bougher et al., 1988, 1990, 1997, 1999, 2002, 2015; Bougher and Borucki, 1994; Brecht and Bougher, 2012; Brecht et al., 2011a, b, 2012, 2021; Gilli et al., 2015). These previous VTGCM modeling efforts for Venus' upper atmosphere (*e.g.,* Parkinson et al., 2021; Bougher et al., 2015; Brecht et al., 2011b; 2012a; Brecht and Ledvina 2020; Brecht and Bougher, 2012b) are directly relevant to analogous exoplanet research performed in this paper.

The VTGCM model domain covers a 5° by 5° latitude-longitude grid, with 69 evenly spaced log-pressure levels in the vertical, extending from ~70 to 300 km (~70 to 200 km) at local noon (midnight). This middle and upper atmosphere domain corresponds to pressures below ~40 mbar. The adopted altitude/pressure range ensures that all the dynamical influences contributing to the NO, $O_2$, and OH nightglow layers can be captured, and the wave processes above the cloud top region can be addressed. Furthermore, the VTGCM can capture the full range of EUV-FUV flux conditions (~1-250 nm) and has been extended to incorporate EUV/FUV/NUV flux conditions for other star types (*e.g.,* Parkinson et al., 2017). Finally, observational comparisons show the VTGCM closely matches Venus' climatological circulation, so we use the present VTGCM dynamical scheme for the 3-D energy balance studies in this paper.

VTGCM groups neutral species into three categories. Major species ($CO_2$, CO, O, $N_2$) influence the atmospheric mean mass, temperature, and global-scale winds; minor species ($O_2$, $N(^4S)$, $N(^2D)$, NO, SO, $SO_2$) which are computed but passive; and 1-D JPL/Caltech KINETICS profiles of specific chemical trace species (*e.g.*, Cl, $Cl_2$, ClCO, ClO, $H_2$, HCl, $HO_2$, $O_3$, OH) from an altitude of ~70 to 110 km (Zhang et al., 2012). See Parkinson et al (2021b) for further details of photochemistry.

Output from the FMS Venus GCM (Lee and Richardson, 2010, 2011) is used to define a lower boundary that is a good representation of the connection between the lower and upper atmospheres (Brecht et al., 2021).

The VTGCM is used in this study to conduct preliminary exoplanet simulations for conditions around the M dwarf star GJ 436 at various star-exoplanet distances (*cf.* Table 1). The self-consistent thermal, dynamical, and chemical equations have been solved to yield steady state solutions.

## *3. VTGCM Simulations: Energy Balance Results and Discussion*

### *3.1 Impact of EUV-UV heating on Energy Balance*

Figure 1 shows a comparison between the solar spectrum (WMO 1985), and a spectrum of the M dwarf GJ 436 from the MUSCLES HST Treasury survey, binned to the JPL/Caltech KINETICS grid. In this figure, the GJ 436 flux is scaled so its total flux equals the total solar flux at 1 AU, $\sim$1360 $Wm^{-2}$. The wavelengths at which photolysis of $CO_2$, $O_2$, $O_3$, and $H_2O_2$ is most efficient are added for comparison (Tian et al., 2014). Recent observations of several planet-hosting M-dwarfs show that most have FUV/NUV flux ratios 1000 times greater than that of the Sun (Tian et al., 2014), and so the relative photolysis rates of the key species (e.g., $CO_2$ and $O_2$) will differ between the two cases. We regard the spectrum of all stellar types to be fixed and do not consider variable spectra in our study.

The M dwarf GJ 436 stellar EUV/FUV fluxes were utilized in conjunction with various star-exoplanet distances for a suite of preliminary/prototype Venus-like exoplanet simulations (see Table 1). The resulting exoplanet geopotential heights, neutral temperatures, heat balances, and cross-terminator zonal winds for the various cases considered thus far are shown in Figures 2, 3, 5, and 6, respectively. Panel (a) in all these figures correspond to VEN1, our baseline Venus case. For all panels (b) – (d), only the planet-star distance is changed as indicated in Table 1 (VEN2 – VEN4 using GJ 436 stellar fluxes, respectively) with no other changes. The VTGCM baseline case run is for solar minimum conditions corresponding to recent *Venus Express* (VEx) observations (Bougher et al., 2015) with upgraded $O_x$, $SO_x$, $HO_x$ chemistry (Mills et al., 2021). These multi-panel figures vividly show the initial comparison of a Venus-like exoplanet versus modern Venus for various planet-star distances from the parent star.

Figure 2 shows that the geopotential height in panels (b) and (c) have some dayside upper atmospheric differences but generally do not vary much from the Venus baseline case shown in panel (a). However, for the closest planet-star distance to GJ 436 (panel (d)) we clearly see a maximum dayside geopotential height of 550 km as compared to a nightside geopotential height of ~220 km. This means that the atmosphere is inflating or bulging due to intense EUV-UV heating on the dayside. Global average heights at a given pressure level are shown on the right-hand side of each plot.

It is seen in Figure 3 that the VTGCM simulations have much faster winds for all cases on the evening terminator (ET) than on the morning terminator (MT), located at 18h and 6h respectively. The maximum MT and ET wind speed are tabulated in Table 2, and we see that the wind speed more than doubled for our VEN4 case and is remarkably fast at the ET. This can be understood by examining Figure 4 which shows the Venus upper atmosphere mean circulation paradigm adopted from Schubert et al. (2007). Previous observations show that Venus' upper atmosphere has two dominating circulation flow patterns (*e.g.,* Bougher et al., 1997, 2006; Lellouch et al., 1997; Schubert et al., 2007; Brecht et al., 2021). The flow known as the retrograde superrotating zonal flow (RSZ), occurs in the region between the surface of the planet to the top of the cloud deck at ∼70 km. This region is dominated by a wind pattern flowing in the direction of the planets spin and is faster than Venus' rotation. The second flow pattern occurs above ∼120 km and is a relatively stable mean subsolar-to-antisolar flow (SS-AS) (Bougher et al., 1997). In the upper atmosphere, Venus has inhomogeneous heating driven mainly by solar radiation (EUV, UV, and IR) thus providing large pressure gradients to generate the dominant SS-AS flow pattern (Dickinson and Ridley, 1977; Schubert et al., 1980; Bougher et al., 1997). The transition region is defined as the altitude range of 70–120 km where the RSZ and SS-AS circulation patterns overlap and are presumed superimposed. This means both flows can be dominant in this altitude region and observations suggest a high degree of variability of these wind components in the transition region (Brecht et al., 2011b, 2021).

The net effect of this general flow pattern paradigm in the upper atmosphere causes (1) a shift in the divergence of the flow from the subsolar point toward the ET, (2) stronger ET winds than those along the MT, and (3) a shift in the convergence of the flow away from midnight and toward the MT (Schubert et al., 2007). This general flow pattern also applies to our exo-Venus case as shown in Figure 3. It is easy to see that the situation varies with altitude, reflecting the changing importance of underlying drivers and possible solar cycle variations.

A comparison of the Venus baseline case (VEN1) and that for GJ 436 fluxes at 0.72 AU (VEN2) shows that: (a) dayside upper thermosphere temperatures ($<10^{-7}$ mbar) (Figure 5(a) vs 5(b)) are up to ~30K cooler for VEN2, and (b) nightside temperatures are similar. The latter indicates the nightside is isolated from the dayside. This isolation is characteristic of a planet with efficient energy loss mechanisms in the upper atmosphere that is rotating sufficiently slowly for it to be close to being tidally locked. Corresponding comparisons among the Venus baseline case (VEN1) and those for GJ 436 fluxes at ~0.38 AU (VEN3) and ~0.175 AU (VEN4) importantly show that: (a) dayside upper thermosphere temperatures (Figure 5(c) vs 5(d)) respectively warm by ~25K and 345K over the Venus baseline case, and (b) nightside temperatures are much the same (isolated from the dayside) and shows that nightside energy is provided by atmospheric circulation rather than radiation. The VEN3 case at the Mercury planet-star distance with the GJ 436 fluxes yields Venus-like current temperatures (*i.e.,* for solar minimum conditions). For all cases thus far, the simulated temperatures are much the same for pressures > $10^{-5}$ mbar (where near IR heating dominates and cooling occurs predominantly via $CO_2$ 15-μm emission). To understand this, we must examine the energy balance for each case. Comparing Figures 3 and 5, we note that at/above about 130 km the slower wind speeds at the MT correlate with a steeper temperature gradient at the MT, contrary to the faster wind speeds and less steep temperature gradient at the ET for all cases VEN1 – VEN4.

In Figure 6, we see that EUV/UV heating and conductive cooling are the dominant heating/cooling terms for pressures < $10^{-7}$ mbar for all cases, roughly balancing each other, and the near IR (NIR) heating is constant for each case. We remark that all cases considered for the Venus baseline case (VEN1) were for solar minimum conditions. If we had considered conditions for solar maximum, such as was the case for Pioneer Venus Orbiter (PVO), NIR heating would have been the same as shown but the EUV/UV heating would be about double in panel (a). The bump in conductive cooling at $\ln(dP_0/dP) = 0$ is positive causing dT/dz to reverse sign and push heat upward instead of downward causing the upper atmosphere to heat up more for larger values.

This is where the $CO_2$ 15-μm band becomes opaque so it can't radiatively cool and where convection becomes important for vertical energy transport. Positive dT/dz means heat is conducted downward (cooling), negative implies heat is conducted upward (heating), and localized heating where dT/dz goes to zero. $CO_2$ 15-μm cooling balances EUV/UV and NIR heating for pressures > $10^{-7}$ mbar with maximum cooling that completely balances the EUV/UV and NIR heating at pressures ~$10^{-5}$ mbar, thus explaining the invariant temperature distribution at pressures of $10^{-5}$ to $10^{-3}$ mbar for all cases. These dayside temperatures ($T_{exo}$ ~ 270 K) would be much warmer than predicted here if the $CO_2$ cooling was not enhanced by collisions of O and $CO_2$. In short, the enhanced non-LTE $CO_2$ 15-μm cooling provides a strong thermostat that helps to regulate dayside temperatures for pressures of $10^{-7}$ to $10^{-2}$ mbar (*e.g.,* Bougher et al., 2002) for the cases shown thus far.

*3.2 Impact of NIR heating on Energy Balance*

The NIR flux should scale with $1/R^2$ just like the UV flux, and Figure 7 illustrates how we parameterize this in our model. Illustrated in this figure are the various NIR heating profiles used for the VTGCM GJ 436 Venus (d = 0.175 AU) NIR heating sensitivity study, obtained by multiplying the standard reference Venus NIR profile at 0.72 AU by various multiplicative factors. The larger the multiplicative factor, the greater the corresponding increase in the NIR heating, parameterizing a range of decreasing Planet-Star distances. Fujii et al. (2017) suggest that the stratospheric dynamical response to increased stellar radiation is similar for stars of different spectral type, but for a warmer star the onset of strong upward motion occurs at a much larger incident flux because less of the instellation is in the NIR where the radiation is more efficiently absorbed by water vapor and clouds. However, a higher temperature star (if it's emitting like a blackbody) will have smaller fraction of flux in the NIR but the absolute flux will be larger because a warmer blackbody has higher absolute emission at all wavelengths. Shown here are the standard reference NIR and 0.7×, 1.3×, 2.0×, and 4.0× the standard reference NIR heating profiles. Figure 8 shows the energy balance for the latter four cases of the NIR sensitivity study for VEN4 in panels (a), (b), (c), and (d), respectively. Kelvin (K)/day units are used throughout both figures. As before (*cf.,* Figure 6), we see that EUV/UV heating and are conductive cooling the dominant heating/cooling terms for pressures < $10^{-7}$ mbar for all cases. The second thing to note in this figure is that moderately large changes of ±30% in the NIR heating profiles corresponding to panels (a) and (b) do not seem to cause much change in the $CO_2$ 15-μm cooling profile, although a minor decrease/increase is seen at $10^{-4}$ mbars for each case compared to Figure 6, panel (d). However, when the NIR heating profile is doubled and quadrupled, the $CO_2$ 15-μm cooling progressively compensates, attempting to maintain the energy balance, as seen in panel (d) where the NIR heating profile maximum matches that of the EUV/UV heating higher in the atmosphere. So, it is seen that the enhanced non-LTE $CO_2$ 15-μm cooling still provides an effective thermostat which helps to regulate dayside temperatures for pressures of $10^{-7}$ to $10^{-2}$ mbar with increasingly very large changes in the NIR heating profile.

One might ask the following: (a) what is the impact on the energy balance of increasing the NIR heating profile as shown in Figure 8, that is to say, just how effective is the enhanced non-LTE $CO_2$ 15-μm cooling, and (b) how much does altering the energy balance by dumping this extra NIR energy throughout the pressure range shown affect key atmospheric parameters for a close-in $CO_2$ dominated atmosphere orbiting an M-dwarf star? Figures 9, 10 and 11 show how changes to the NIR heating/energy balance affect the geopotential height, zonal wind, and temperature results, respectively. In each of these figures, panels (a), (b), (c), and (d) show the resultant VEN4 plots with 0.7×, 1.3×, 2.0×, and 4.0× the standard reference NIR heating, respectively.

Figure 9 shows that the dayside local noon geopotential height maximum progressively bulges with increasing NIR heating, ballooning from ~300 ± 20 km for the VEN1, VEN2, and VEN3 cases (*cf.,* Table 2) out to 630 km for the VEN4 1.0× case and 760 km for the VEN4 4.0× case (*cf.,* Table 4). Typical nightside local midnight geopotential height maximums are ~210 km for the VEN1, VEN2, and VEN3 cases, yielding an dayside noon to nightside midnight elongation ratio of ~3:2. Correspondingly, from Figure 9 it is seen that the nightside local midnight geopotential height maximum is ~220 km and ~250 km for the VEN4 2.0× case and the VEN4 4.0× case, respectively, giving a remarkable approximate dayside noon to nightside midnight elongation ratio of ~3:1 for these cases.

Figure 10 represents the mean zonal wind plots to test the impact of the NIR heating sensitivity study. As was evident in Figure 3, the VTGCM simulations have much faster winds for all cases on the evening terminator (ET) than on the morning terminator (MT), located at 18h and 6h respectively. The maximum MT and ET wind speed are tabulated in Table 3, and we see that both the MT and ET wind speeds increased by ~45% for our VEN4 4.0× case over the VEN4 1.0× case and is remarkably fast at the ET at 650 m/s. The effect of quadrupling the NIR heating profile compared to doubling, Figure 7, shows the maximum wind speed increases are much greater for this case than those for changes occurring due to successive changes from 1.3x and 2.0x the standard reference NIR profile. The progressive increase of the top of the nightside midnight geopotential height and the dramatic increase in and distribution of the MT and ET zonal winds shown in Figures 9 and 10, panels (a) – (d) is further proof that nightside energy is provided by atmospheric circulation rather than radiation.

From Figure 11, our best estimate thus far is that 0.7× NIR cools and 1.3× NIR warms dayside temperatures near the relative pressure given by $\ln(P_0/P$ (left axis of the plots in the figure) of -6.0 to -5.0 by about ±10K. Previously shown in Figure 6 in the relative pressure region at altitudes below $\ln(P_0/P) = -10$, there was little or no change in the temperature structure, and comparing Figures 6 panel (d) and 11 panels (a) and (b), we see that this corresponds to a temperature of ~160K for 0.7×, 1.0×, and 1.3× NIR, respectively. However, we see in Figure 11 panel (d) for this region that there is a net difference in temperature of about 80K over the value of 160K to about 240K for the VEN4 4× NIR case.

Most of the action is higher up in the lower thermosphere and upper mesosphere as expected. For the region where $\ln(P_0/P) \geq 0$, changes ±30% (*cf.,* Figure 11, panels (a) and (b)) in the NIR heating, nightside peak temperatures don't show much change and the dayside peak temperatures warm (cool) by about only +15K (-15K) over the standard reference NIR heating profile case shown in Figure 6 panel (d). Thus, from our simulations we see these moderate changes in the NIR heating do not result in large changes in temperature over the standard reference NIR heating case. However, doubling and quadrupling the NIR heating profile produces significant changes in the entire temperature structure of the upper and middle atmosphere. Nightside peak temperatures increase by as much as 30K and the dayside peak temperatures warm by a dramatic 50K and 130K for the doubled and quadrupled NIR heating profiles over the standard reference NIR heating profile case shown in Figure 6 panel (d). At the $\ln(P_0/P) = 3\text{-}5$ level (just above the dT/dz inflection point) we have a doubling of absolute temperature between panels (b) and (d) in this region. It is to be noted that even an exospheric temperature in excess of 500K, slightly less than twice the Venusian one, means that the Lambda parameter (Gronoff et al., 2020) of atmospheric escape is about 11. This implies that we are far from hydrodynamic escape (it would require values below 2.5) and assures a low atmospheric escape rate under quiet stellar conditions.

Table 3 summarizes some key results of the sensitivity of changes in the NIR heating profile on the energy balance discussed above for our close-in $CO_2$ dominated exoplanetary atmosphere orbiting the M-dwarf star, GJ 436. Remarkably, we see that even in the extreme cases shown, the enhanced non-LTE $CO_2$ 15-μm cooling still provides a strong thermostat that helps to regulate dayside temperatures for pressures of $10^{-7}$ to $10^{-2}$ mbar.

## *4. Conclusions*

These are the first 3-D simulations with self-consistent dynamics, energetics, and photochemistry focused on the observable upper atmosphere of Venus-like exoplanets. We have modified the VTGCM to include Venus-like exoplanet conditions by: (a) extending the stellar EUV-FUV-NUV flux bins to 250 nm, and (b) applying the planet-star scaling to all stellar fluxes (EUV/FUV/NUV). We also utilize validated radiative heating and cooling schemes to generate GCM thermal diagnostics. We have compared GCM thermal balances for an M-dwarf star type (*i.e.*, terms will include stellar EUV-FUV-NUV heating, $CO_2$ 15-μm (thermal infrared) cooling, molecular thermal conduction, impacts of both horizontal and vertical winds) to quantify how the dayside heat budget changes with star type (our Sun and GJ 436) and planet-star distance.

Our calculations show that with decreasing planet-star distance from GJ 436, the dayside temperatures progressively increase, and morning and evening terminator winds become stronger. Moderate changes in NIR heating do not impact the energy balance but very large changes in NIR heating do, altering the geopotential height, zonal wind speed, and temperature structure of the middle and upper atmosphere significantly. Most importantly, while temperatures are increasing with increasing EUV-UV and NIR heating, non-LTE $CO_2$ 15-μm cooling provides a strong thermostat that helps to regulate dayside temperatures for all cases. These dayside temperatures would be much warmer than predicted here if the $CO_2$ cooling were not enhanced by collisions of O and $CO_2$. Interestingly, nightside temperatures remain relatively constant for all cases where the NIR heating is kept constant but has a little variability in response to changes in the NIR heating profile (especially in the upper atmosphere).

This study shows that Venus-like planets are likely to be more stable against atmospheric escape than Earth-like planets, since the EUV-XUV fluxes have lower influence on their, very low, exospheric temperatures. The $CO_2$ thermostat regulates temperatures so that atmospheric escape is less likely, and a $CO_2$ dominated atmosphere on exoplanets around M-dwarf stars could be preferred and possibly detectable. While this does not prevent atmospheric escape through non-thermal processes, which will require further studies, it supports the opinion that we are more likely to find Venus-like exoplanets around M-dwarfs (Kane et al., 2021) (*i.e.,* $CO_2$ atmospheres). JWST is slated to specifically look for such $CO_2$ terrestrial like planets and it is therefore to be expected that it will detect such atmospheres. Future research will include photochemical studies of key species of astrobiological interest (*viz.,* $CO_2$, CO, $CH_4$, $H_2O$, $H_2$, $O_2$, $O_3$, $NH_3$, $H_2O_2$, and $H_2C_2$) that combined with these results, could assist in providing very good ways of indicating the existence of such Venus-like atmospheres in the targets of future high-resolution telescopes.

Temperatures affect the photodissociation cross-sections, particularly the long wavelength dependence of the cross section. Extending the VTGCM solar flux out to 400 nm will be examined as a part of future work and is particularly important for $CO_2$ since otherwise, one would miss the high temperature hot bands for wavelengths greater than 250 nm. This points to the importance of proper temperatures for photochemical calculations that have temperature dependent reaction rates to calculate volume mixing ratios of key species of astrobiological significance.

This paper focuses on the energy balance, temperature structure, and dynamics of an exo-Venus a parallel paper is to follow regarding photochemistry and airglow.

### *5. Availability of data and materials*

All datasets and software used for supporting the conclusions of this article are available from the University of Michigan Deep Blue Data Repository website at https://doi.org/10.7302/vtaz-tw05 .

**Acknowledgements**. CDP would like to acknowledge to Danica Adams who assisted with the plotting of the figures in this paper. YLY was supported in part by an NAI Virtual Planetary Laboratory grant from the University of Washington. DJA was supported by a NASA FINESST Fellowship.

| Case | Stellar Flux | Planet-Star distance (AU) | Comments |
|---|---|---|---|
| VEN1 | Sun ($s_{min}$) | 0.72 | Venus Baseline |
| VEN2 | Scaled GJ 436 | 0.72 | Fluxes change only |
| VEN3 | Scaled GJ 436 | 0.38 | Fluxes + Star-Planet distance |
| VEN4 | Scaled GJ 436 | 0.175 | Fluxes + Star-Planet distance |

**Table 1:** Various Test Simulations

| Case | Max Geopotential Height Near Equator | Dayside $T_{exo}$ Noon near Equator | Max MT vs ET Winds near Equator | Max Dayside $CO_2$ 15-μm Cooling and EUV/UV/IR Heat Balance$^{\dagger}$ |
|---|---|---|---|---|
| VEN1 | 320 km | 246 K | -150/+275 m/s | 3000 K/day |
| VEN2 | 280 km | 210 K | -125/+250 m/s | 2500 K/day |
| VEN3 | 320 km | 270 K | -150/+275 m/s | 3000 K/day |
| VEN4 | 630 km | 620 K | -350/+450 m/s | 12,000 K/day |

**Table 2:** Various parameters from Figures 2, 3, 5, 6. $^{\dagger}$All days are terrestrial days.

| Case | NIR Multiplicative Factor | Max Geopotential Height Near Equator | Dayside $T_{exo}$ Noon near Equator | Max MT vs ET Winds near Equator | Max Dayside $CO_2$ 15-μm Cooling and EUV/UV/IR Heat Balance$^{\dagger}$ |
|---|---|---|---|---|---|
| VEN4 | 0.7 | 620 km | 610 K | -300/+400 m/s | ~11,000 K/day |
| VEN4 | 1.3 | 640 km | 640 K | -350/+450 m/s | ~12,000 K/day |
| VEN4 | 2.0 | 680 km | 670 K | -400/+500 m/s | ~12,500 K/day |
| VEN4 | 4.0 | 760 km | 750 K | -500/+650 m/s | ~15,000 K/day |

**Table 3:** Near Infrared sensitivity study: Various parameters from Figures 8-11. $^{\dagger}$All days are terrestrial days.

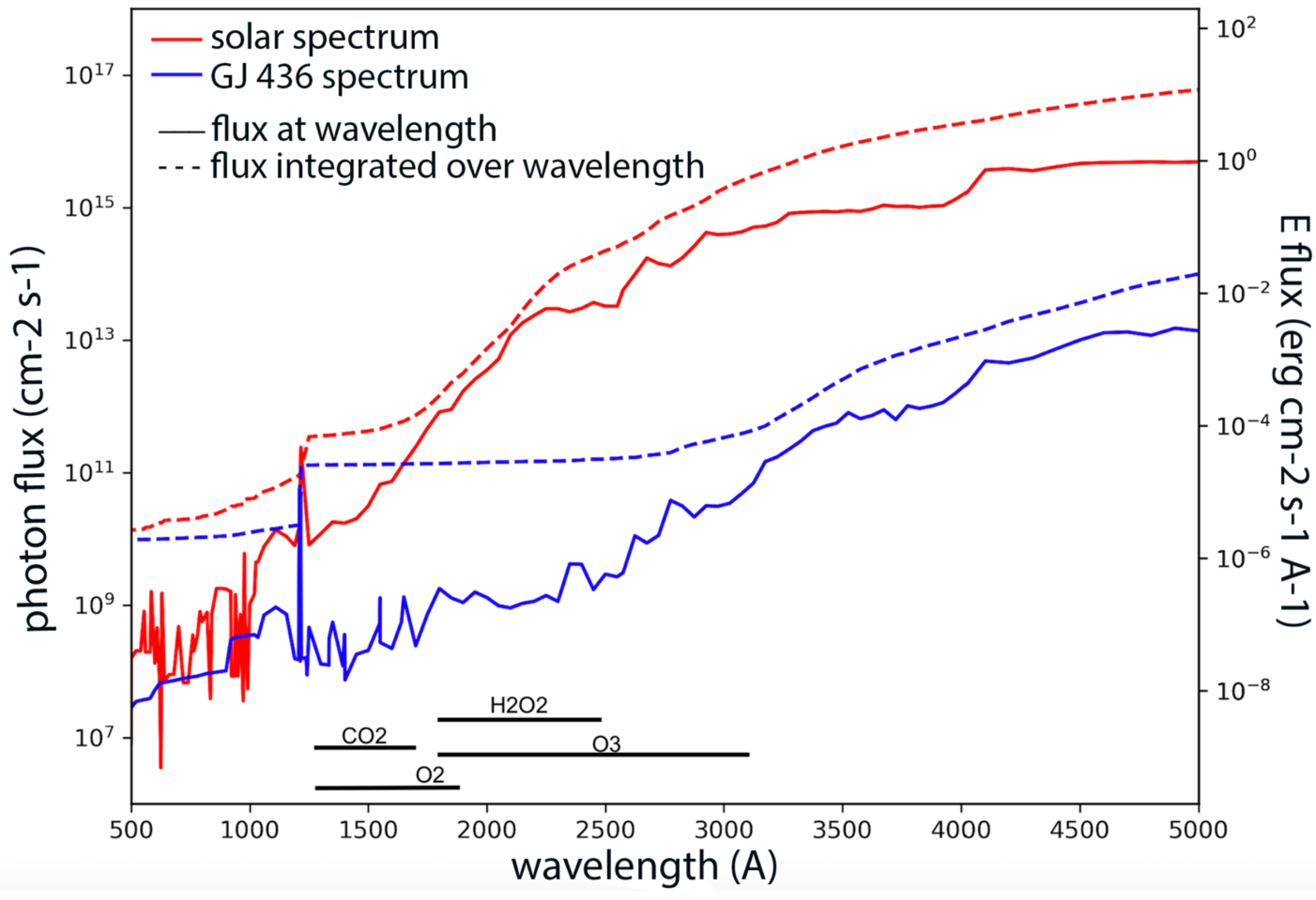

Figure 1: The spectra of the Sun (WMO 1985; red) and GJ 436 photon flux, integrated photon flux, and energy flux (from the MUSCLES HST Treasury survey, binned to the JPL/Caltech KINETICS grid) and scaled such that the total flux of each is identical to the total flux received at 1 AU (~1360 $Wm^{-2}$) (Gao et al., 2015; Tian et al., 2014). The wavelengths at which photolysis of $CO_2$, $O_2$, $O_3$, and $H_2O_2$ is most efficient are added for comparison (Tian et al., 2014). GJ 436 is brighter in the IR and contributes to normalization to the total flux of 1360 $Wm^{-2}$.

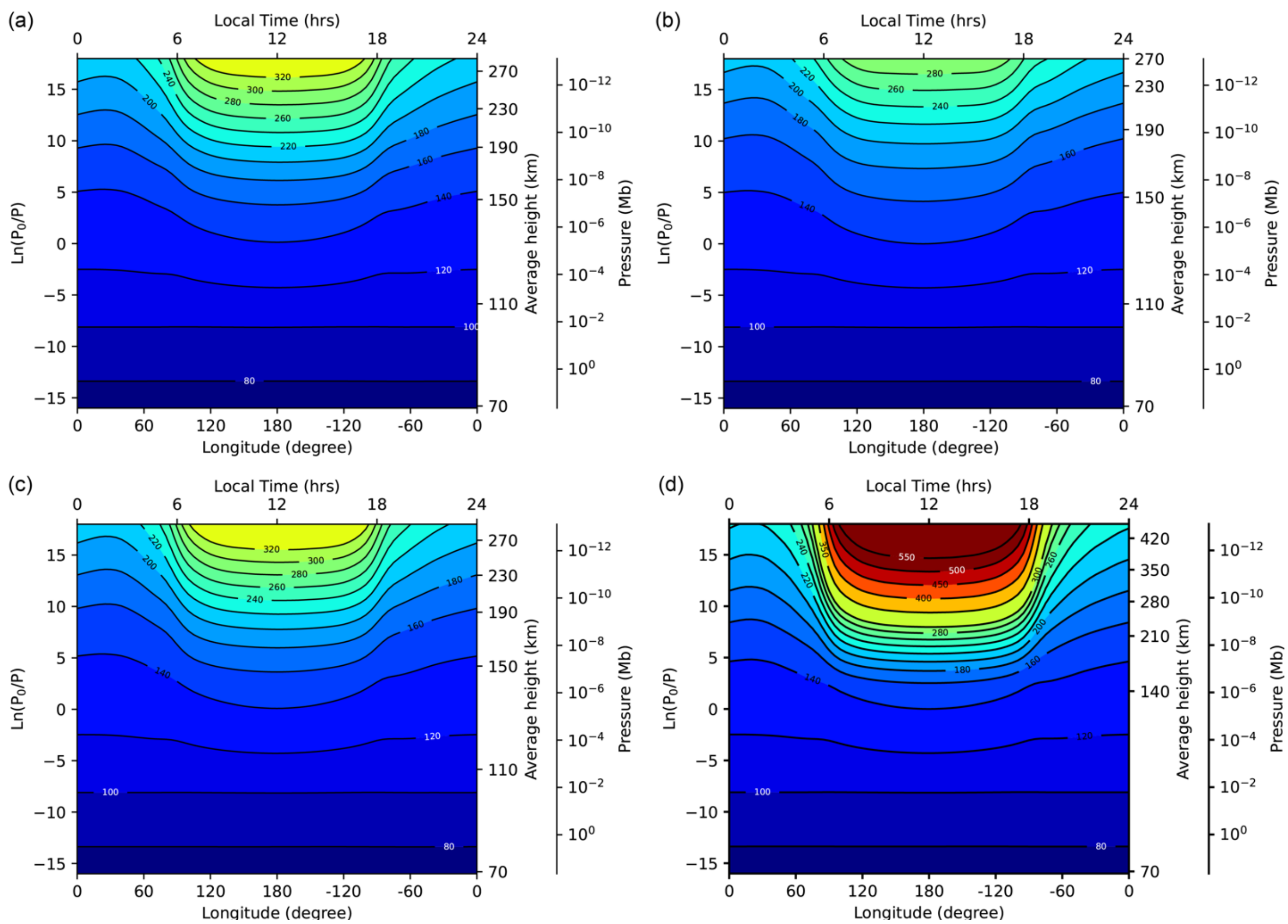

Figure 2: Geopotential Height. Panel (a) represents present Venus VTGCM base case for solar minimum conditions (validated against Venus Express observations); panel (b) VTGCM GJ 436 Venus at d = 0.72 AU (present Venus orbital distance); panel (c) VTGCM GJ 436 Venus at d = 0.387 AU (present Mercury orbital distance); panel (d) VTGCM GJ 436 Venus at d = 0.175 AU. All panels are at 2.5°N latitude (equatorial).

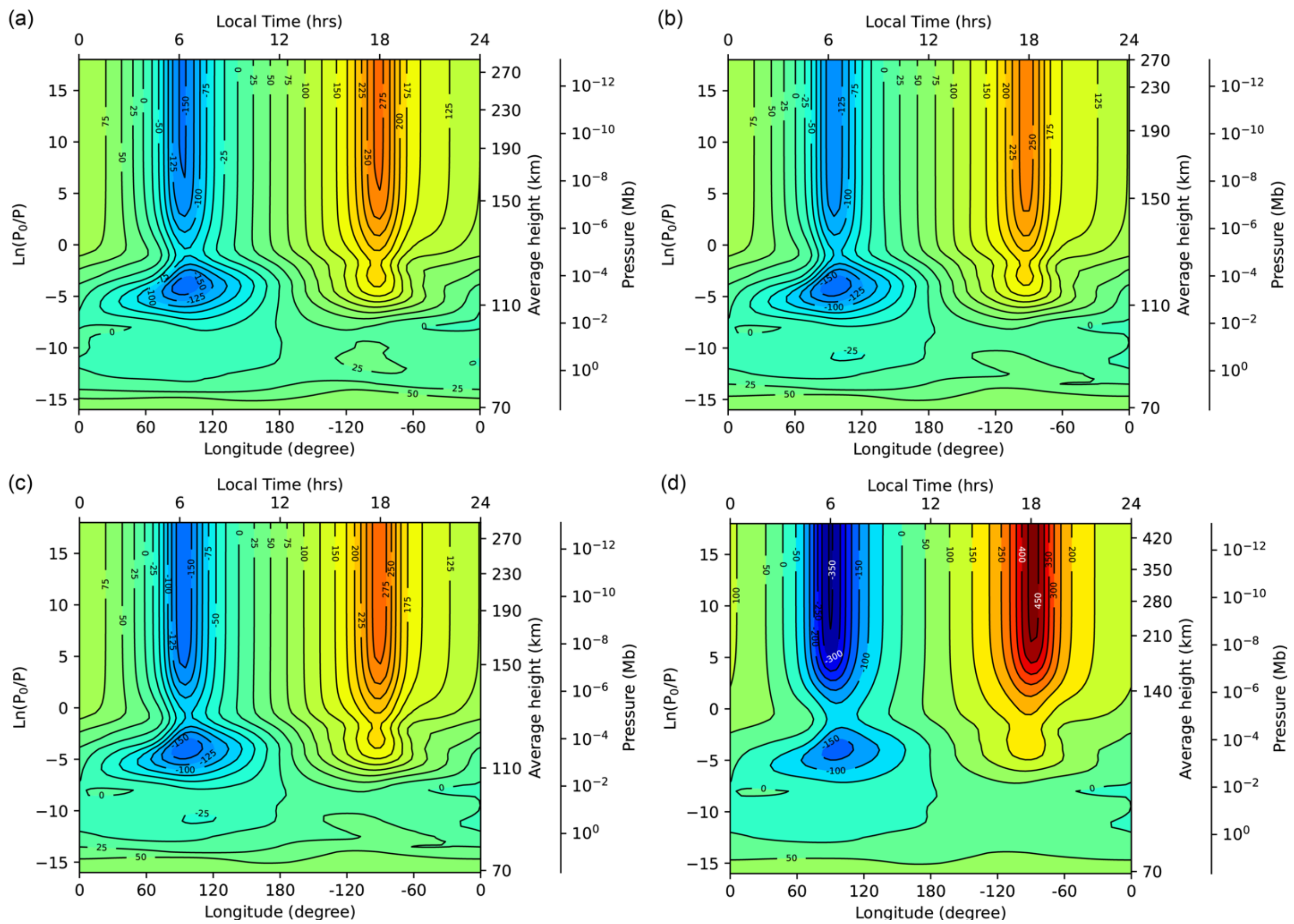

Figure 3: **u**(z) Zonal Wind. Panel (a) represents present Venus VTGCM base case for solar minimum conditions (validated against Venus Express observations); panel (b) VTGCM GJ 436 Venus at d = 0.72 AU (present Venus orbital distance); panel (c) VTGCM GJ 436 Venus at d = 0.387 AU (present Mercury orbital distance); panel (d) VTGCM GJ 436 Venus at d = 0.175 AU. All panels are at 2.5°N latitude (equatorial). $m \cdot s^{-1}$ units are used throughout.

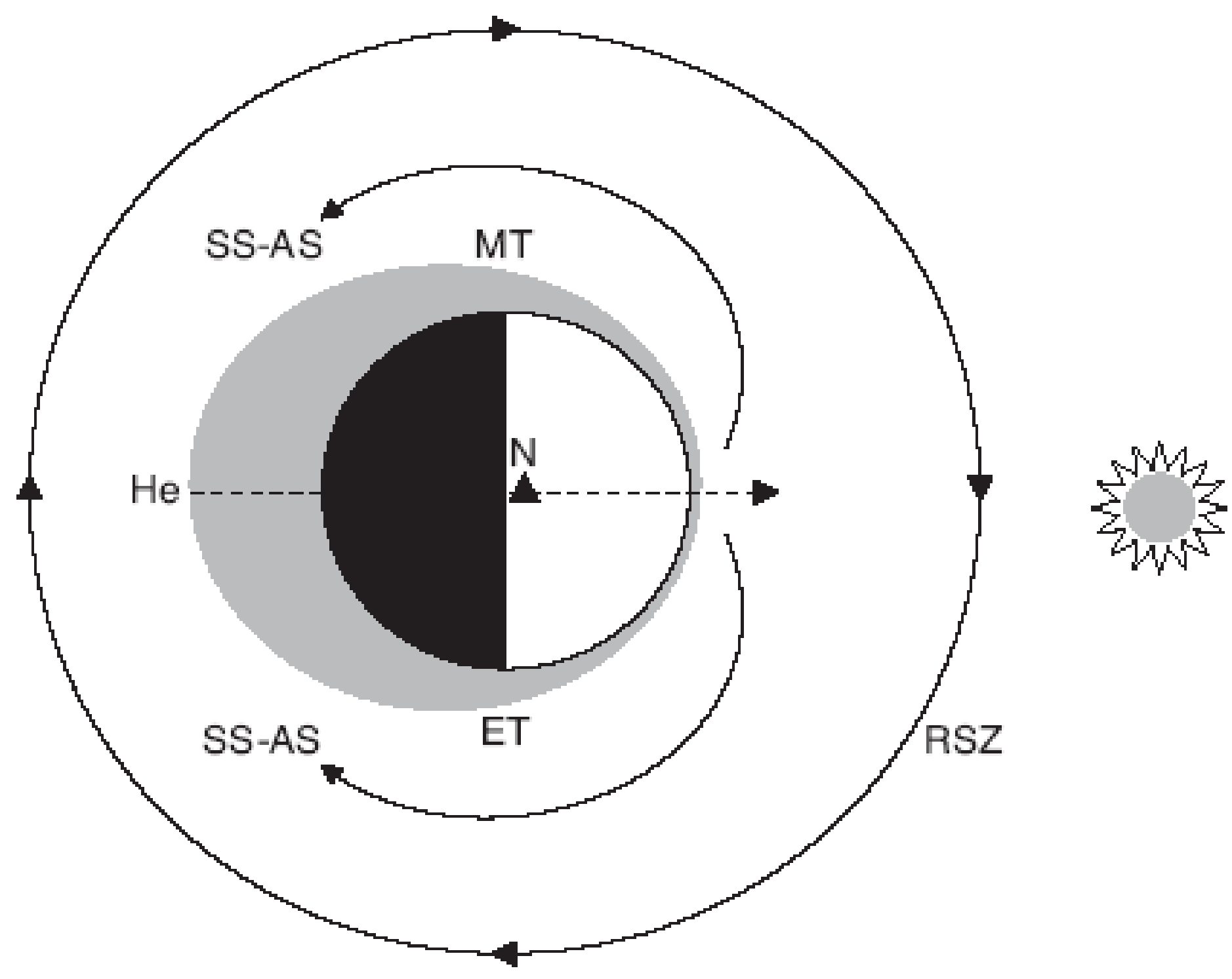

Figure 4: Simple Venus upper atmosphere circulation paradigm. Here, MT = morning terminator; ET = evening terminator; SS-AS = stable subsolar to antisolar circulation cell driven by NIR and EUV heating; RSZ = retrograde superrotating zonal flow that seems to vary greatly over time; and He = helium tracer bulge. Adopted from Schubert et al. (2007).

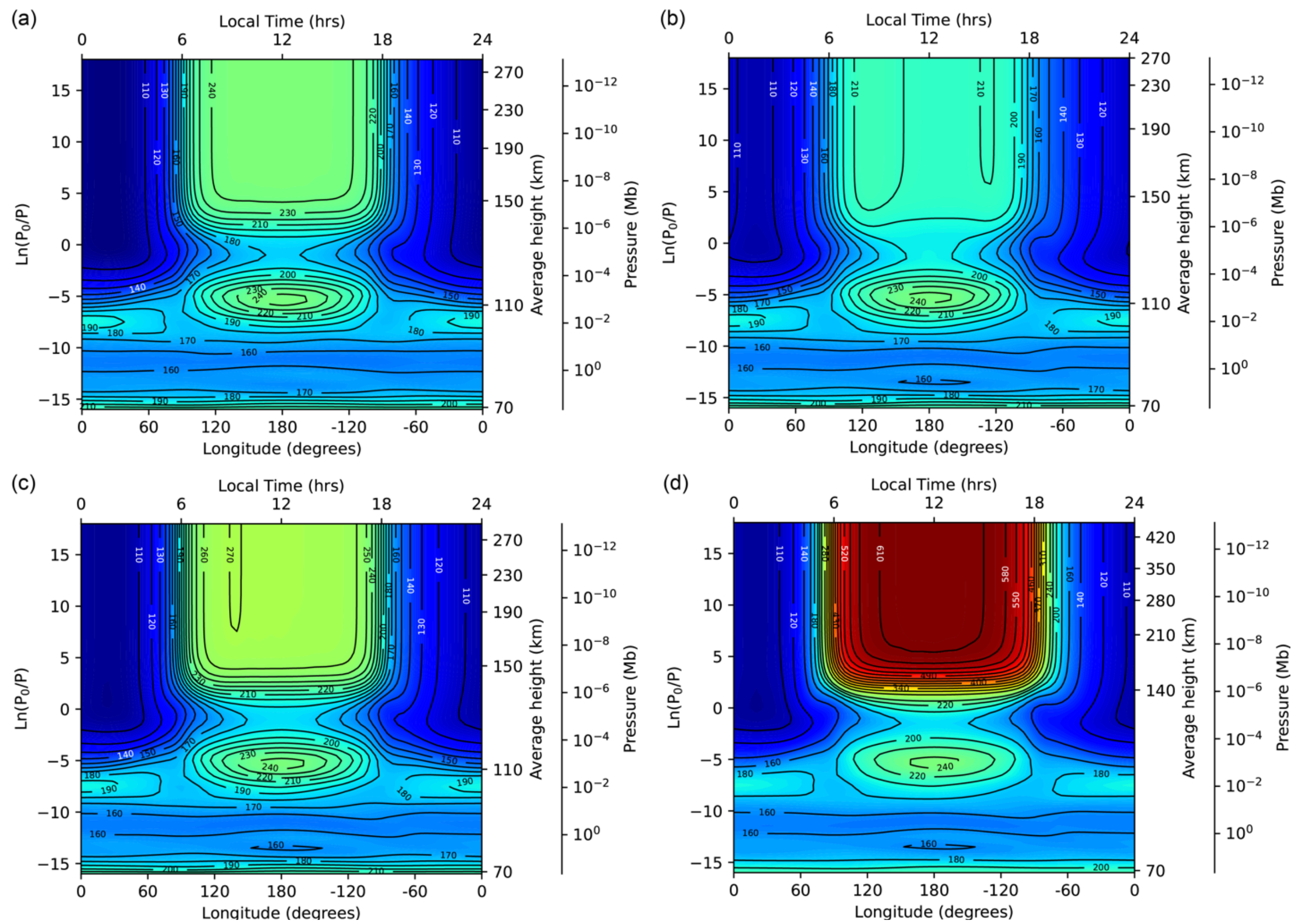

Figure 5: Neutral Temperature for all cases. Panel (a) represents present Venus VTGCM base case for solar minimum conditions (validated against Venus Express observations); panel (b) VTGCM GJ 436 Venus at d = 0.72 AU (present Venus orbital distance); panel (c) VTGCM GJ 436 Venus at d = 0.387 AU (present Mercury orbital distance); panel (d) VTGCM GJ 436 Venus at d = 0.175 AU. All panels are at 2.5°N latitude (equatorial). Kelvin (K) units are used throughout.

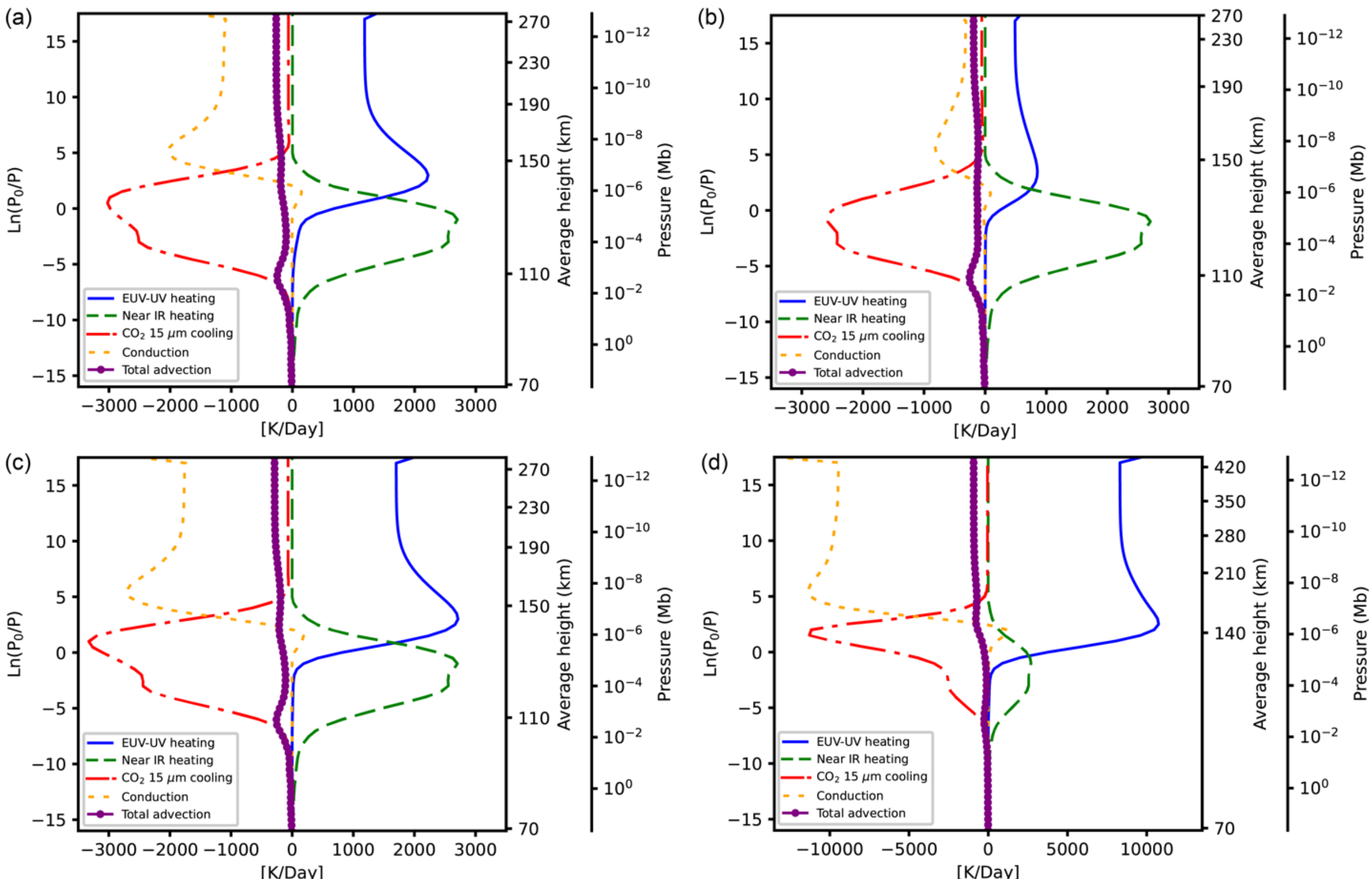

Figure 6. Energy balance analysis with Venus NIR heating constant for all cases. Panel (a) represents Venus VTGCM base case for solar minimum conditions (validated against Venus Express observations); panel (b) VTGCM GJ 436 Venus at d = 0.72 AU (Venus orbital distance); panel (c) VTGCM GJ 436 Venus at d = 0.387 AU (Mercury orbital distance); panel (d) VTGCM GJ 436 Venus at d = 0.175 AU. All panels are at 2.5°N latitude, local time noon, and for pressures ranging from ~50-$10^{-12}$ mbar. Kelvin (K) /day units are used throughout.

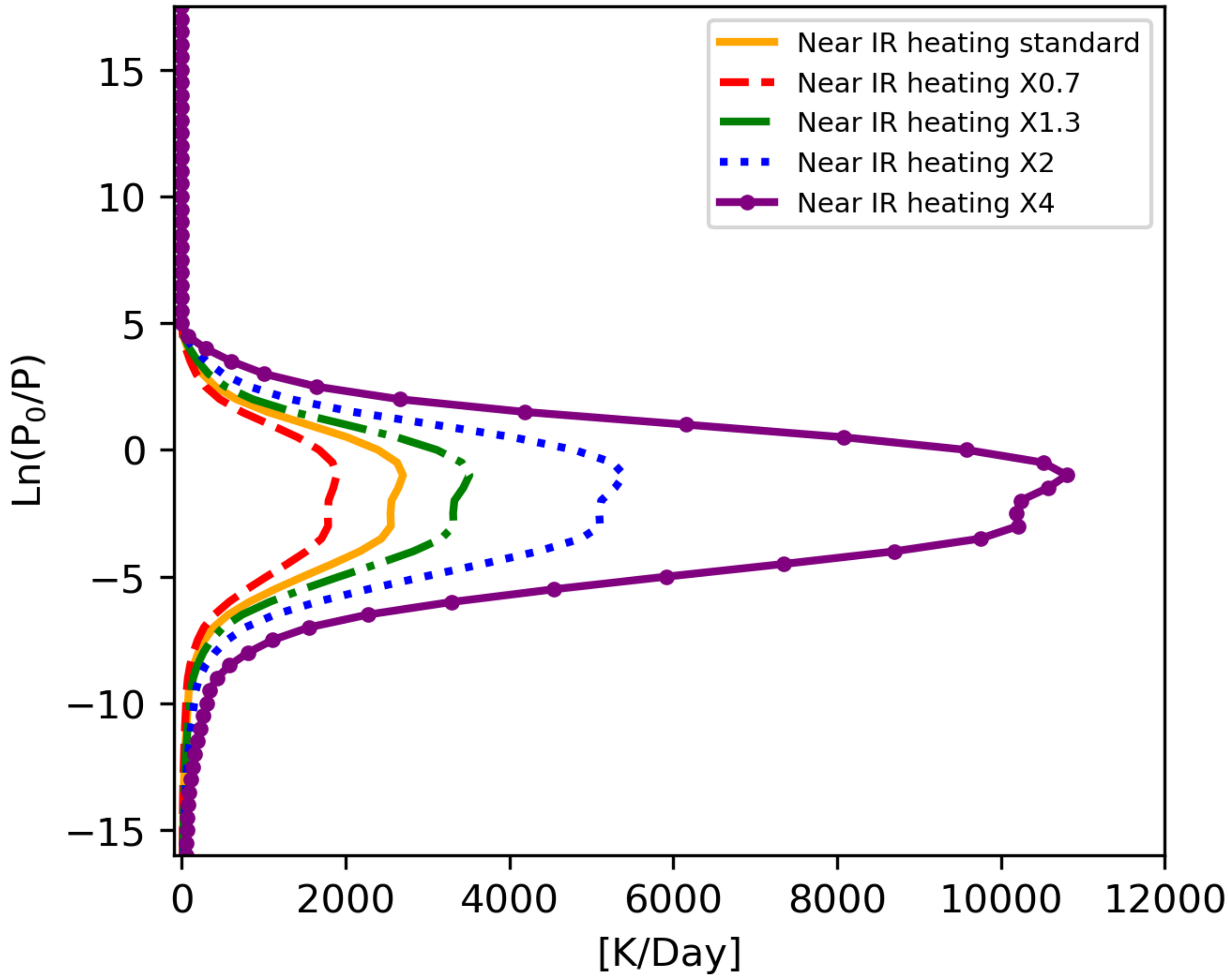

Figure 7. Changes to the standard reference NIR (std. ref. NIR) heating profile for the energy balance sensitivity study. Shown are 0.7× std. ref. NIR (red dashed), standard reference NIR (gold solid), 1.3× std. ref. NIR (green dot dashed), 2.0× std. ref. NIR (blue dotted), 4.0× std. ref. NIR (purple solid with dots) heating profiles. Kelvin (K)/day units are used for pressures ranging from ~50-$10^{-12}$ mbar.

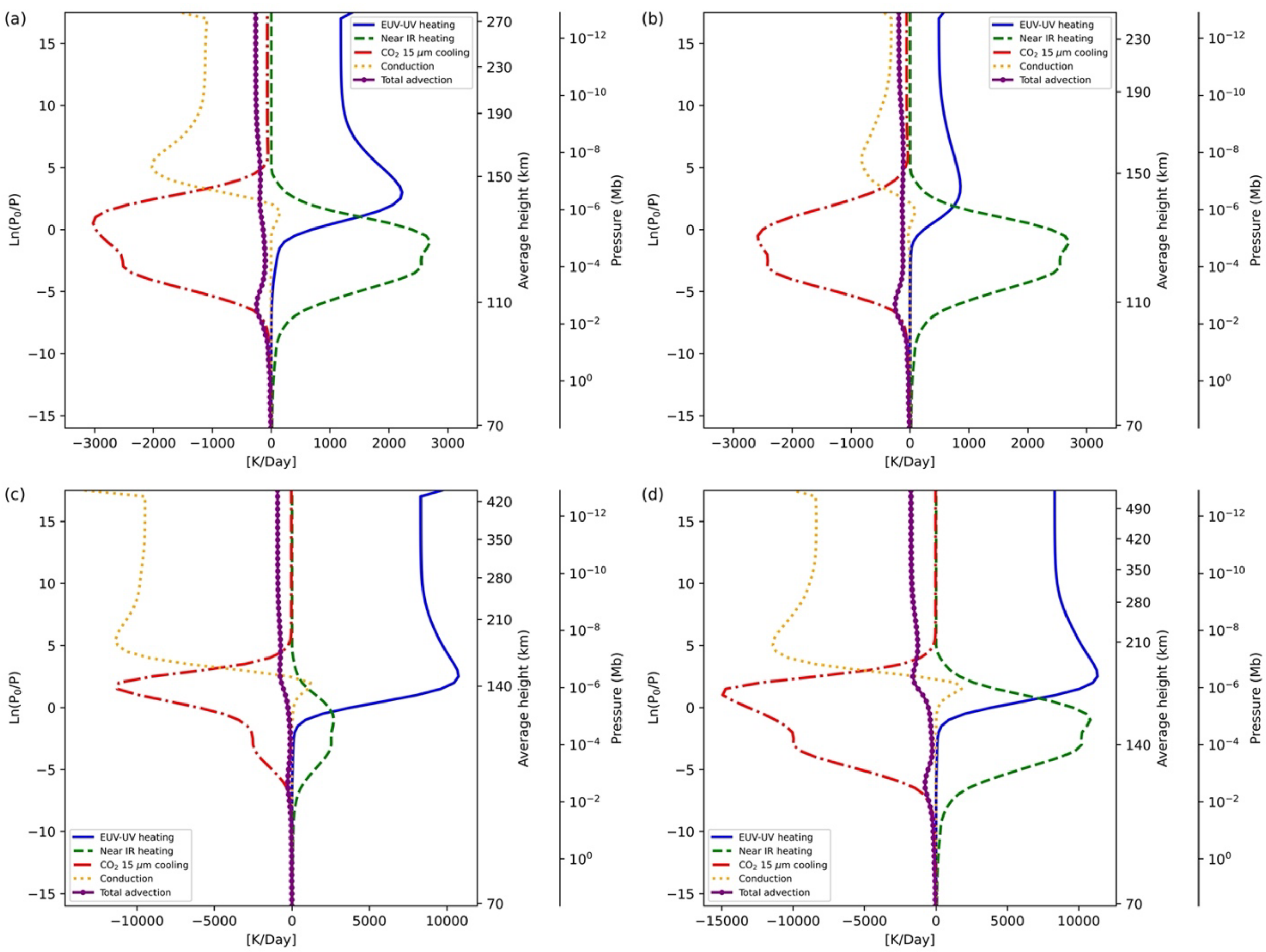

Figure 8. Energy balance sensitivity study analysis by varying the std. ref. NIR heating. Comparison is for the VEN4 case (VTGCM GJ 436 Venus at d = 0.175 AU). Panels (a) through (d) respectively show the resultant VEN4 energy balance plots for the 0.7× std. ref. NIR, 1.3× std. ref., 2.0× std. ref. NIR, 4.0× std. ref. NIR heating profiles shown in Figure 6. All panels are at 2.5°N latitude (equatorial) and Kelvin per day (K/day) units are used throughout. All panels are for pressures ranging from ~50-$10^{-12}$ mbar.

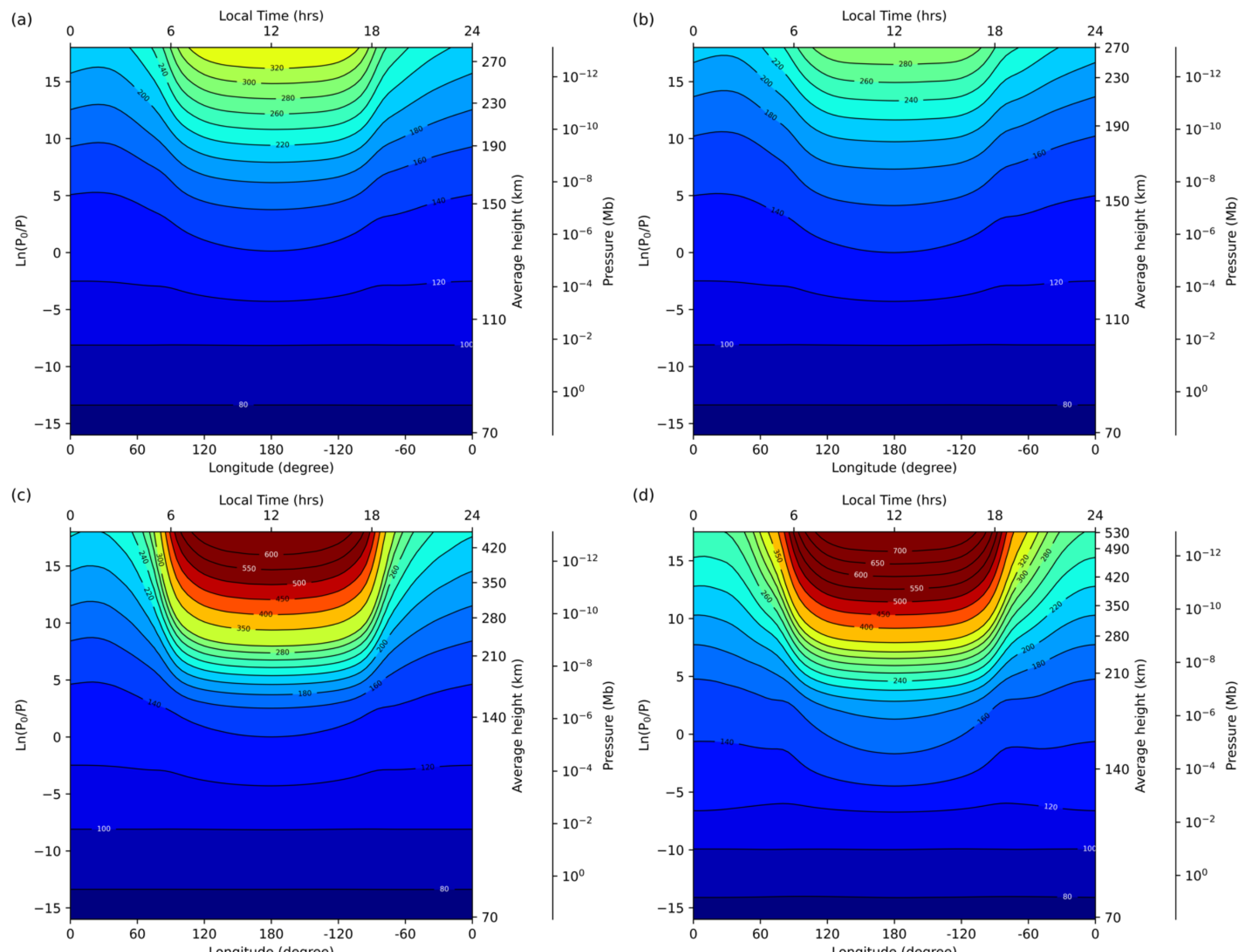

Figure 9. Geopotential Height results for the energy balance sensitivity study. Comparison for the VEN4 case (VTGCM GJ 436 Venus at d = 0.175 AU). Panels (a) through (d) respectively show the resultant VEN4 geopotential plots for the 0.7× std. ref. NIR, 1.3× std. ref., 2.0× std. ref. NIR, 4.0× std. ref. NIR heating profiles shown in Figure 6. All panels are at 2.5°N latitude (equatorial) and for pressures ranging from ~50-$10^{-12}$ mbar.

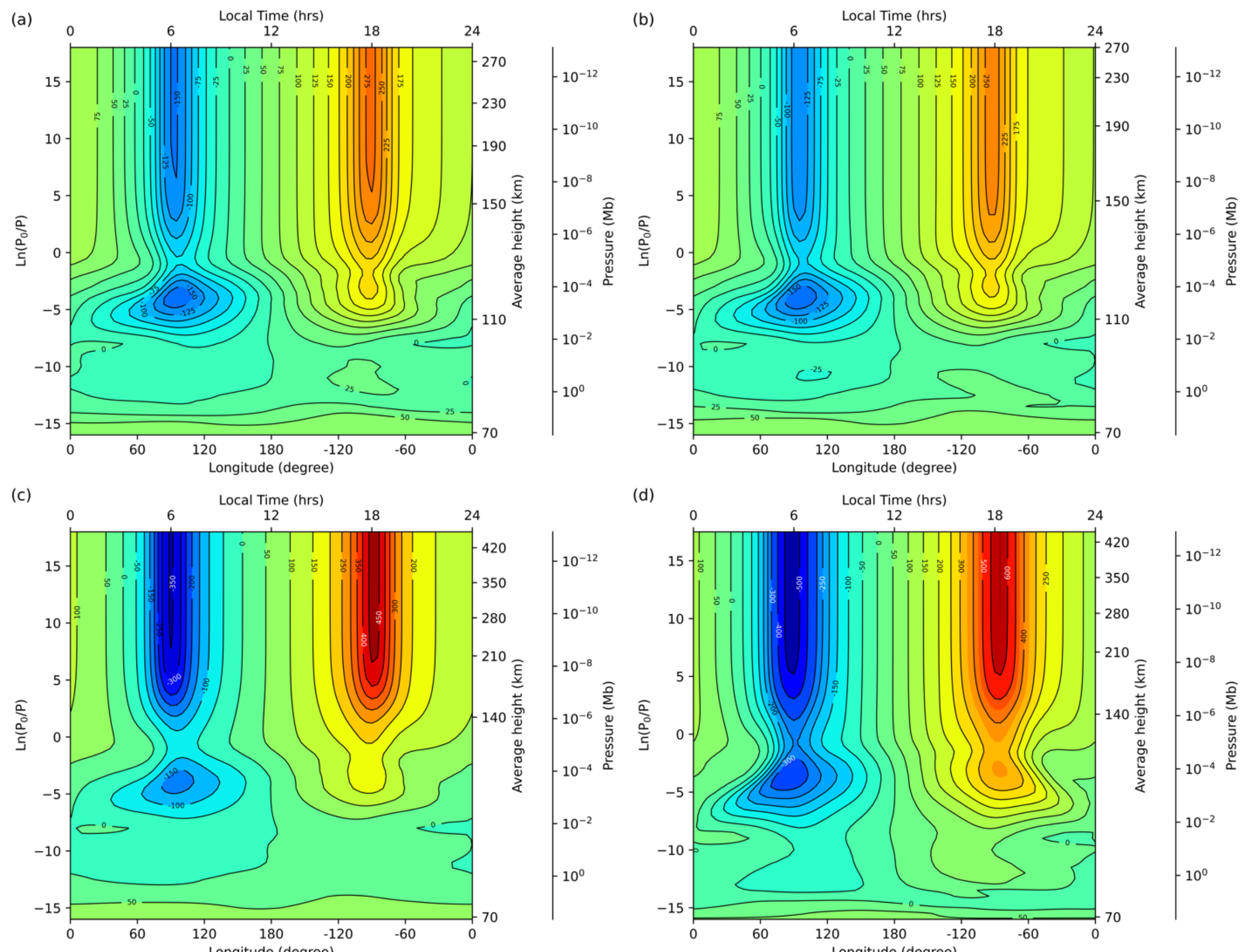

Figure 10. **u**(z) Zonal Wind results for the energy balance sensitivity study. Comparison for the VEN4 case (VTGCM GJ 436 Venus at d = 0.175 AU). Panels (a) through (d) respectively show the resultant VEN4 **u**(z) zonal wind plots for the 0.7× std. ref. NIR, 1.3× std. ref., 2.0× std. ref. NIR, 4.0× std. ref. NIR heating profiles shown in Figure 6. All panels are at 2.5°N latitude (equatorial) and m/s units are used throughout. All panels are for pressures ranging from ~50-$10^{-12}$ mbar.

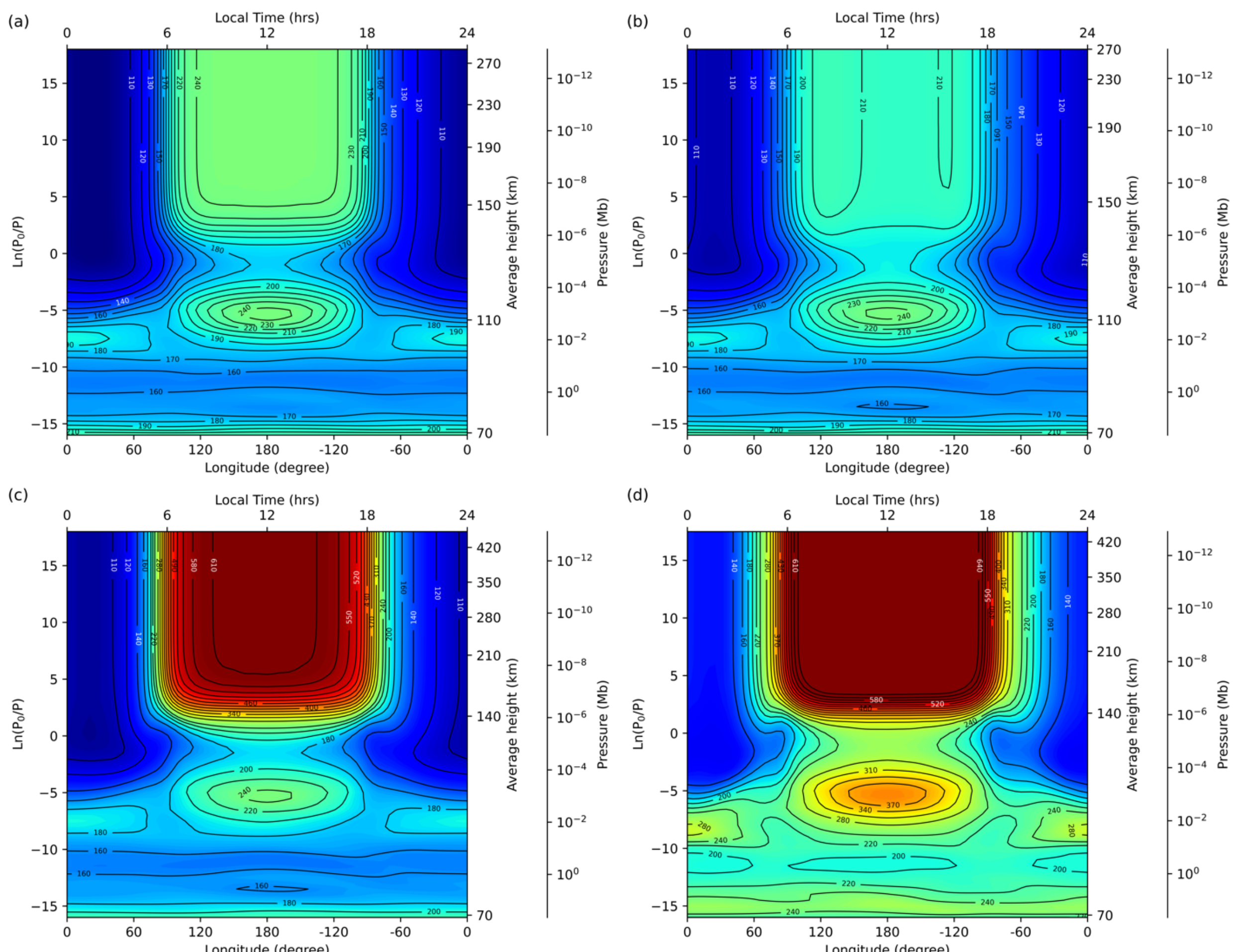

Figure 11. Neutral Temperature results for the energy balance sensitivity study. Comparison for the VEN4 case (VTGCM GJ 436 Venus at d = 0.175 AU). Panels (a) through (d) respectively show the resultant VEN4 temperature plots for the 0.7× std. ref. NIR, 1.3× std. ref., 2.0× std. ref. NIR, 4.0× std. ref. NIR heating profiles shown in Figure 6. All panels are at 2.5°N latitude (equatorial) and Kelvin (K) units are used throughout. All panels are for pressures ranging from ~50-$10^{-12}$ mbar.

## References

Airapetian, V. S., A. Glocer, G. Gronoff, E. Hébrard, and W. Danchi.. "Prebiotic chemistry and atmospheric warming of early Earth by an active young Sun." *Nature Geoscience*, 9 (6): 452-455, 2016, [10.1038/ngeo2719]

Airapetian, V. S., R. Barnes, O. Cohen, et al.. "Impact of space weather on climate and habitability of terrestrial-type exoplanets." *International Journal of Astrobiology*, **19 (2):** 136-194, 2019, [10.1017/s1473550419000132]

Allen, M., Yung, Y. L., Waters, J. W., Vertical transport and photochemistry in the terrestrial mesosphere and lower thermospehere. *Journal of Geophysical Research,* 86, 3617–3627, 1981.

Arney, G. N., The K Dwarf Advantage for Biosignatures on Directly Imaged Exoplanets", *Ap. J.*, 837:L7, 2019, https://doi.org/10.3847/2041-8213/ab0651.

Arney G. N., Meadows V. S., Domagal-Goldman S. D., et al., *Ap. J.,* 836, 49, 2017.

Arney G., Domagal-Goldman S. D., Meadows V. S. et al., *Astrobiology,* 16, 873, 2016.

Benneke, B., & Seager, S., *Ap. J.*, 753, 100, 2012.

Belyaev, D., et al. Vertical profiling of $SO_2$ above Venus' clouds by means of SPICAV/SOIR occultations. *Icarus,* 217, 740-751, 2012.

Berner, R. A., & Canfield, D. E., Am.J.Sci., 289, 333, 1989.

Beust, H. et al. Dynamical evolution of the Gliese 436 planetary system. Kozai migration as a potential source for Gliese 436b's eccentricity. *Astron. Astrophys.,* **545**, A88, 2012.

Bougher, S. W., and W. J. Borucki, Venus O2 visible and IR nightglow: Implications for lower thermosphere dynamics and chemistry, *J. Geophys. Res.*, *99*, 3759-3776, 1994.

Bougher, S. W., et al., Venus mesosphere and thermosphere: II. Global circulation, temperature, and density variations, *Icarus*, *68*, 284-312, 1986.

Bougher, S. W., R. E. Dickinson, E. C. Ridley, and R. G. Roble, Venus mesosphere and thermosphere III. Three-dimensional general circulation with coupled dynamics and composition, *Icarus*, *73*, 545-573, 1988.

Bougher, S. W., J. C. Gérard, A. I. F. Stewart, and C. G. Fesen, The Venus nitric oxide night airglow: Calculations based on the Venus thermospheric general circulation model, *J. Geophys. Res.*, *95*, 6271-6285, 1990.

Bougher, S. W., D. M. Hunten, and R. G. Roble, CO2 cooling in terrestrial planet thermospheres, *J. Geophys. Res*., *99*, 14609-14622, 1994.

Bougher, S. W., M. J. Alexander, and H. G. Mayr, Upper atmosphere dynamics: Global circulation and gravity waves, in *Venus II*, pp. 259-292, U. of Arizona Press, 1997.

Bougher, S. W., S. Engel, R. G. Roble, and B. Foster, Comparative terrestrial planet thermospheres: 2. Solar cycle variation of global structure and winds at equinox, *J. Geophys. Res.*, *104*, 16591-16611, 1999.

Bougher, S. W., R. G. Roble, and T. J. Fuller-Rowell, Simulations of the upper atmospheres of the terrestrial planets, in AGU Monograph: *Comparative Aeronomy in the Solar System,* Eds. M. Mendillo, A. F. Nagy, and J. H. Waite, 2002.

Bougher, S. W., S. Rafkin, and P. Drossart, Dynamics of the Venus upper atmosphere: Outstanding problems and new constraints expected from Venus Express, *Planet. Space Sci.*, *54*, 1371-1380, doi:10.1016/j.pss.2006.04.23, 2006.

Bougher, S. W., J.M. Bell, J.R. Murphy, M.A. Lopez-Valverde, P.G. Withers *Geophys. Res. Lett.* **33**, DOI:10.1029/2005GL024059, 2006.

Bougher, S. W., P.-L. Blelly, M. Combi, J. L. Fox, I. Mueller-Wodarg, A. Ridley, and R. G. Roble, Neutral Upper Atmosphere and Ionosphere Modeling, *Space Sci. Reviews*, 139, 107-141, doi:10.1007/s11214-008-9401-9, 2008.

Bougher, S. W., McDunn, T. M., Zoldak, K. A., Forbes, J. M., *GRL*, 36, L05201, 2009.

Bougher, S. W., D. Pawlowski, J. M. Bell, S. Nelli, T. McDunn, J. R. Murphy, M. Chizek, and A. Ridley, Mars Global Ionosphere-Thermosphere Model (MGITM): Solar cycle, seasonal, and diurnal variations of the Mars upper atmosphere, *J. Geophys. Res. Planets*, 120, 311–342, 2015, doi:10.1002/2014JE004715.

Bougher, S. W., A. S. Brecht, R. Schulte, J.-L. Fischer, C. D. Parkinson, A. Mahieux, V.Wilquet, A.-C. Vandaele, Upper Atmosphere Temperature Structure at the Venusian Terminators: A Comparison of SOIR and VTGCM Results, *Planet. Space Sciences*, 2015, http://dx.doi.org/10.1016/j.pss.2015.01.012.

Bougher, S. W., D. Pawlowski, J. M. Bell, S. Nelli, T. McDunn, J. R. Murphy, M. Chizek, and A. Ridley, *JGR: Planets* 120, 311, doi:10.1002/2014JE004715, 2015a.

Bougher, S.W. et al., *Science*, 350, #6261 doi:10.1126/science.aad0459, 2015b.

Bougher, S.W. et al., Variability of Mars Thermosphere Neutral Structure from MAVEN Deep Dip Observations: NGIMS Comparisons with Global Models, *AGU Fall 2015*, Abst. # P21A-2064, 2015c.

Bougher, S. W. et al., *JGR: Space Physics*, 122, 1296-1313. doi:10.1002/2016JA023454, 2017a.

Bougher, S. W., et al.. Mars Dayside Thermospheric Composition and Temperatures from the NGIMS MAVEN Instrument: Implications for Thermal Balances, *DPS 2017*, Abst# 510.5, 2017b.

Bougher, S. W., D. A. Brain, J. L. Fox, F. Gonzalez-Galindo, C. Simon-Wedlund, and P. G. Withers. Chapter 14: Upper Atmosphere and Ionosphere, in *The Atmosphere and Climate of Mars*, ed. B. Haberle, M. Smith, T. Clancy, F. Forget, R. Zurek, Cambridge University Press, 2017c, doi:10.1017/9781107016187.

Bourrier, V., Lovis, C., Beust, H. *et al.* Orbital misalignment of the Neptune-mass exoplanet GJ 436b with the spin of its cool star. *Nature,* **553,** 477–480, 2018. https://doi.org/10.1038/nature24677

Brecht, A. S., Tracing the Dynamics in Venus' Upper Atmosphere, Dissertation, University of Michigan, 2011a.

Brecht, A. S., S. W. Bougher, J. C. Gérard, C. D. Parkinson, S. Rafkin, B. Foster, Understanding the Variability of Nightside Temperatures, NO UV and $O_2$ IR Nightglow Emissions in the Venus Upper Atmosphere, *J. Geophys. Res.,* 116, E08004, doi:10.1029/2010JE003770, 2011b.

Brecht, A. S., S. W. Bougher, J. C. Gérard, L. Soret, Atomic Oxygen Distributions in the Venus Thermosphere: Comparisons Between Venus Express Observations and Global Model Simulations, *Icarus*, 117, 759-766, 2012a.

Brecht, A. S. and S. W. Bougher, Dayside Thermal Structure of Venus' Upper Atmosphere Characterized by a Global Model, *J. Geophys. Res.*, 117, E8, doi:10.1029/2012JE004079, 2012b.

Brecht, A. S., Bougher, S. W., Shields, D., and Liu, H. L., Planetary-Scale Wave Impacts on the Venusian Upper Mesosphere and Lower Thermosphere. Journal. *Geophys. Res*., 126, doi:10.1029/2020JE006587, 2021.

Brecht, S. H., & Ledvina, S. A.. An explanation of the nightside ionospheric structure of Venus. *JGR: Space Physics*, 126, e2020JA027779, 2021. https://doi.org/10.1029/2020JA027779

Broeg, C., Benz, W., Thomas, N., Cheops Team. The CHEOPS mission. *Contrib. Astron. Obs. Skaln. Pleso*., 43:498, 2014.

Burkholder, J. B., S. P. Sander, J. Abbatt, J. R. Barker, R. E. Huie, C. E. Kolb, M. J. Kurylo, V. L. Orkin, D. M. Wilmouth, and P. H. Wine "Chemical Kinetics and Photochemical Data for Use in Atmospheric Studies, Evaluation No. 18," JPL Publication 15-10, Jet Propulsion Laboratory, Pasadena, 2015, http://jpldataeval.jpl.nasa.gov.

Chameides, W. L., & Walker, J. C., *OLEB*, 11, 291, 1981.

Catling, D., & Kasting, J. F., *Atmospheric Evolution on Inhabited and Lifeless Worlds* (Cambridge: Cambridge Univ. Press), 2017.

Clancy, R.T., Sandor, B.J., Moriarty-Schieven, G., Circulation of the Venus Upper Mesosphere/Lower Thermosphere: Doppler Wind Measurements from 2001-2009 Inferior Conjunction, Submillimeter CO Absorption Line Observations. *Icarus*, doi: 10.1016/j.icarus.2011.05.021, 2012b.

Collet, A., C. Cox, and J. C. Gérard, Two-dimensional time-dependent model of the transport of minor species in the Venus night side upper atmosphere, *Planet. Space Sci,* 58 (14), 1857-1867, 2010.

Coughlin, J.L., Stringfellow, G.S., Becker, A.C., López-Morales, M., Mezzalira, F., & Krajci, T.. Parameter Variations in the Transits of Gliese 436b. *Ap. J. Lett.*, doi: 10.1086/595822, 2008.

Crisp, D., et al., Ground-based near-infrared observations of the Venus nightside: 1.27-micron O2(1-Delta) airglow from the upper atmosphere, *J. Geophys. Res.*, *101*, 4577-4594, 1996.

Crisp, D., Radiative forcing of the Venus mesosphere I: Solar fluxes and heating rates. *Icarus* 67, 484–514, 1986.

Des Marais, D. J., Harwit, M., Jucks, K., Kasting, J. F., Lunine, J. I., Lin, D., Seager, S., Schneider, J., Traub, W., & Woolf, N., Remote Sensing of Planetary Properties and Biosignatures on Extrasolar Terrestrial Planets, *Astrobiology*, 2, 153-181, 2002.

Deming, D., and S. Seager, Illusion and Reality in the Atmospheres of Exoplanets, *J. Geophys. Res.,* arXiv:1701.00493, 2017.

DeMore, W. B., Sander, S. P., Golden, D. M., Hampson, R. F., Kurylo, M. J., Howard, C. J., Ravishankara, A. R., Kolb, C. E., Molina, M. J.,. Chemical kinetics and photochemical data for use in stratospheric modeling: Evaluation number 12. *JPL Publication* 97-4, Jet Propulsion Laboratory, Pasadena CA, 1997.

Demory, Brice-Olivier & Gillon, M. & Waelkens, C. & Queloz, Didier & Udry, S.. (2009). GJ 436c? The contribution of transit timings. *Proceedings of The International Astronomical Union*. 253. 424-427. 10.1017/S1743921308026835.

Domagal-Goldman, S. D., Segura, A., Claire, M. W., et al., *Ap. J.*, 792, 90, 2014.

Dickinson, R. E., and E. C. Ridley, Venus mesosphere and thermosphere temperature structure. II - Day-night variations, Icarus, 30, 163–178, doi:10.1016/0019-1035(77)90130-0, 1977.

Ehrenreich, D., Bourrier, V., Wheatley, P. *et al.* A giant comet-like cloud of hydrogen escaping the warm Neptune-mass exoplanet GJ 436b. *Nature,* **522,** 459–461. https://doi.org/10.1038/nature14501, 2015.

Fauchez T. J., Turbet M., Villanueva G. L. et al., *Ap. J.,* **887** 194, 2019.

Fox, J. L., and K. Y. Sung, Solar activity variations of the Venus thermosphere/ionosphere, *J. Geophys. Res.*, *106* (A10), 21305-21336, 2001.

Fridlund, et al. *Space Sci. Rev*., 2016.

Fujii, Y., A.D. Del Genio, and D.S. Amundsen, NIR-driven moist upper atmospheres of synchronously rotating temperate terrestrial exoplanets. *Astrophys. J.*, **848**, no. 2, 100, 2017, doi:10.3847/1538-4357/aa8955.

Gao, P., Hu, R., Robinson, T. D., Li, C., & Yung, Y. L., *Ap. J.*, 806, 249, 2015.

Garcia, R. F., P. Drossart, G. Piccioni, M. López-Valverde, and G. Occhipinti, Gravity waves in the upper atmosphere of Venus revealed by $CO_2$ nonlocal thermodynamic equilibrium emissions, *J. Geophys. Res.*, 114(E00B32), doi: 10.1029/2008JE003073, 2009.

Gonzalez-Galindo, F. et al., *J. Geophys, Res*., **118,** 2105-2123, 2013.

Gillon, M., Triaud, A. H. M. J., Demory, B.-O., et al., *Nature*, 542, 456, 2017.

Grimm, S. L., Demory, B.-O., Gillon, M., et al., *A&A*, 613, A68, 2018.

Gronoff, G., …, C. D. Parkinson, and other authors. Atmospheric Escape Processes and Planetary Atmospheric Evolution, JGR: Space Physics, 125, 8, DOI10.1029/2019.JA027639, 2020.

Haberle R. M. et al., Int'l Workshop: Mars Atmosphere Modeling and Observations, Abstract, Granada, Spain, Jan. 13-15, 2003.

Harman, C. E., F. Felton, R. Hu, S. D. Domagal-Goldman, A. Segura, F. Tian, and J.F. Kasting, Abiotic O2 Levels on Planets around F, G, K, and M Stars: Effects of Lightning-produced Catalysts in Eliminating Oxygen False Positives, *Astrophys. J.*, 866, 56, https://doi.org/10.3847/1538-4357/aadd9b , 2018.

Hayworth, B., Gronoff, G., Airapetian V., Hegyi B., and Kasting, J., Impact of an Active Young Sun on the Hadean Environment in revision, *PNAS*, 2022.

Hu et al., *Ap. J.*, 888, 122, 2020.

Hu, R., Seager, S., & Bains, W., *Ap. J.*, 761, 166, 2012.

Jessup et al., , Coordinated Hubble Space Telescope and Venus Express Observations of Venus' upper cloud deck, *Icarus,* doi:10.1016/j.icarus.2015.05.027, 2015.

Kane, S.R., Kopparapu, R.K., Domagal-Goldman, S.D.. *Astrophys. J.*, 794, L5, 2014.

Kane, S.R., Ceja, A.Y., Way, M.J., Quintana, E.V.. *Astrophys. J.*, 869, 46, 2018.

Kane, S.R., et al. (2019). JGR: Planets, 124, 2015, 2019.

Kane, S.R., et al. (2021). *JGR: Planets*, 126, 06643 Madden, J.H., Kaltenegger, L. (2018). *Astrobiology*, 18, 1559, 2021.

Kasting, J. F., Evolution of Oxygen and ozone in the Earth's Atmosphere, PhD Dissertation, Univ. Michigan, 1979.

Kasting, J. F., Runaway and moist greenhouse atopsheres and the evolution of Earth and Venus, *Icarus*, 74, 472, 1988.

Kasting, J. F., Bolide impacts and the oxidation state of carbon in the Earth's early atmosphere, *Origins of Life*, 20, 199, 1990.

Kasting, J. F., Eggler, D. H., & Raeburn, S. P., Mantle redox evolution and the case for a reduced Achean atmosphere, *J. Geol*., 101, 245, 1993a.

Kasting, J. F., Whitmire, D. P., & Reynolds, R. T., Habitable zones around main sequence stars, *Icarus*, 101, 108, 1993b.

Kiang, N. Y., Siefert, J., & Blankenship, R. E., *Astrobiology*, 7, 222, 2007.

Komacek, T., Showman, A., and Tan, X. Atmospheric Circulation of Hot Jupiters: Dayside-Nightside Temperature Differences. II. Comparison with Observations. *The Astrophysical Journal.* 835. 10.3847/1538-4357/835/2/198, 2016.

Kopparapu, R. K., Ramirez, R., Kasting, J. F., et al., *Ap. J.*, 765, 131, 2013.

Krissansen-Totton, J., Fortney, J. J., Nimmo, F., & Wogan, N. Oxygen false positives on habitable zone planets around sun-like stars. *AGU Advances*, 2, e2020AV000294. https://doi.org/10.1029/2020AV000294, 2021.

Krasnopolsky, V. A. & Parshev, V. A., *Nature* 292, 610–613, 1981.

Kump, L. R., *Nature*, 451, 277, 2008.

Léger, A., Fontecave, M., Labeyrie, A., et al., *Astrobiology*, 11, 335, 2011.

Lellouch, E., T. Clancy, D. Crisp, A. J. Kliore, D. Titov, and S. W. Bougher, Monitoring of mesospheric structure and dynamics, in Venus II: Geology, Geophysics, Atmosphere, and Solar Wind Environment, edited by S. W. Bougher, D. M. Hunten, and R. J. Phillips, pp. 295–324, Univ. of Ariz. Press, Tucson, 1997.

Lenton, T. M., GCBio, 7, 613, 2001.
Li, C., Le, T., Zhang, X., and Yung, Y. L., "A High-performance Atmospheric Radiation Package: with Applications to the Radiative Energy Budgets of Giant Planets", *JQSRT*, 217, 353-362, 2018.
Lincowski, A. P., Meadows, V. S., Crisp, D., Robinson, T. D., Luger, R., Lustig-Yaeger, J., and Arney, G. N.: Evolved Climates and Observational Discriminants for the TRAPPIST-1 Planetary System, *Ap. J.*, 867, 1-34, doi: 10.3847/1538-4357/aae36a, 2018.
Lippincott, E. R., Eck, R. V., Dayhoff, M. O., & Sagan, C., *Ap. J.*, 147, 753, 1967.
Lovelock, J. E., *Nature*, 207, 568, 1965.
Luger, R., & Barnes, R., *Astrobiology*, 15, 119, 2015.
Lyons, T. W., Reinhard, C. T., & Planavsky, N. J., *Nature*, 506, 307, 2014.
MacKevoy and Tirion, Cambridge Double Star Atlas, 2nd ed., 2018.
McDunn, T. L., Bougher, S. W., Murphy, J., Smith, M. D., Forget, F., Bertaux, J.-L., Montmessin, F., *Icarus* **206**, 5, 2010.
Madhusudhan, N., Agundez, M., Moses, J., and Hu, Y., Exoplanetary Atmospheres – Chemistry, Formation, Conditions, and Habitability, *Space Sci. Rev.*, December; 205(1): 285–348. 2016, doi:10.1007/s11214-016-0254-3.
Mahieux, A., A. C. Vandaele, R. Drummond, S. Robert, V. Wilquet, A. Fedorova, and J. L. Bertaux, Densities and temperatures in the Venus mesosphere and lower thermosphere retrieved from SOIR onboard Venus Express: Retrieval technique, *J. Geophys. Res*., 115, E12014, 2010.
Mahieux, A., A. C. Vandaele, S. Robert, V. Wilquet, R. Drummond, F. Montmessin, and J. L. Bertaux, Densities and temperatures in the Venus mesosphere and lower thermosphere retrieved from SOIR onboard Venus Express: Carbon dioxide measurements at the Venus terminator, *J. Geophys. Res.,* 117, E07001, doi:10.1029/2012JE004058, 2012.
Marcq, E., F. Mills, C. D. Parkinson, B. Sandor, A. C. Vandaele, Composition and Chemistry of the Neutral Atmosphere of Venus, Space Science Reviews 214(1), DOI10.1007/s11214-017-0438-5, 2017.
Marcq, E., Belyaev, D., Montmessin, F., Fedorova, A., Bertaux, J.L., Vandaele, A.C., Neefs, E., An investigation of the $SO_2$ content of the venusian mesosphere using SPICAV-UV in nadir mode, *Icarus,* 211, 58-69, 2011.
Marcq, E., J.-L. Bertaux, F. Montmessin, and D. Belyaev, Variations of sulphur dioxide at the cloud top of Venus's dynamic atmosphere, *Nature Geo.,* doi:10.1038/NGEO1650, 2012.
Meadows, V. S., *Astrobiology*, 17, 1022, 2017.
Meadows, V. S., Reinhard, C. T., Arney, G. N., et al., arXiv:1705.07560, 2017.
Meadows, V. S., Arney, G. N., Schwieterman, E. W., Lustig-Yaeger, J., Lincowski, A. P., Robinson, T., Domagal-Goldman, S. D., Deitrick, D., Barnes, R. K., Fleming, D. P., Luger, R., Driscoll, P. E., Quinn, T. R., and Crisp, D.: The Habitability of Proxima Centauri b: Environmental States and Observational Discriminants, *Astrobiology* 18, 2018. doi: 10.1089/ast.2016.1589.
Medvedev, A. S., E. Yiğit, P. Hartogh, and E. Becker, Influence of gravity waves on the Martian atmosphere: General circulation modeling, *J. Geophys. Res*., 116, E10004, doi:10.1029/2011JE003848, 2011.
Mickol, R. L., Farris, H. N., Kohler, E., Chevrier, V., Kral, T. A., and Lacy, C.. A SIMPLE ONE-DIMENSIONAL RADIATIVE-CONVECTIVE ATMOSPHERE MODEL FOR USE

WITH EXTRASOLAR ATMOSPHERES. 46th Lunar and Planetary Science Conference., 2015.

Miles, B. E., & Shkolnik, E. L., arXiv:1705.03583, 2017.

Mills F.P., J.I. Moses, P. Gao, and S.-M. Tsai, The diversity of planetary atmospheric chemistry: Lessons and challenges from our solar system and extrasolar planets, *Space Sci. Rev*. 217, 43, doi: 10.1007/s11214-021-00810-1, 2021a.

Mills, F.P., et al., Proceedings of the Venera-D Venus Modelling Workshop October 5-7, 2017, 59-62, Moscow 2018.

Molina, M. and F. J. Rowland, Stratospheric sink for chlorofluoromethanes: chlorine atom-catalysed destruction of ozone, *Nature*, 249, 810-812, 1974.

Montabone, L., F. Forget, E. Millour, R. J. Wilson, S. R. Lewis, B. Cantor, D. Kass, A. Kleinbohl, M. T. Lemmon, M. D. Smith, M. J. Wolff, Eight-year climatology of dust optical depth on Mars, *Icarus*, 251, 65-95, http://dx.doi.org/10.1016/j.icarus.2014.12.034, 2015.

Nair, H., Allen, M., Anbar, A. D., et al., *Icarus*, 111, 124, 1994.

Navarro-Gonzalez, R., McKay, C. P., & Mvondo, D. N., *Nature*, 412, 61, 2001.

Nna Mvondo, D., Navarro-González, R., McKay, C. P., Coll, P., & Raulin, F., *AdSpR*, 27, 217, 2001.

Parkinson, C. D., A. Brecht, S. W. Bougher, F. Mills, and Y. L. Yung, Photochemical Distribution of Venusian Sulphur and Halogen Species*, B.A.A.S. Trans*., DPS, Fall Meet. Suppl., Abstract # 10.04, 2010b.

Parkinson, C., Yung, Y., Esposito, L., Gao, P., Bougher, S.W., Hirtzig, M., 2014. Photochemical Control of the Distribution of Venusian Water and Comparison to Venus Express SOIR Observations. *Planet. Space Sci*., doi:10.1016/j.pss.2015.02.015, 2015a.

Parkinson, C. D., et al., Analysis of Venus Express optical extinction due to aerosols in the upper haze of Venus, *Planet. Space Sci.*, doi:10.1016/j.pss.2015.01.023, 2015b.

Parkinson, C. D., et al., On Understanding the Nature and Variation of the Venusian Middle Atmosphere Via Observations and Numerical Modeling of Key Tracer Species, Venera-D Modeling Workshop, Moscow, Russia, 2017.

Parkinson, C. D., and Bougher, S. W., Venus-like Exoplanet Simulations: Temperature, Wind, Energy Balance, and Species Distribution Datasets Using the VTGCM, University of Michigan Deep Blue Data Repository, [Dataset], https://deepblue.lib.umich.edu/data/concern/data_sets/d504rk80j, https://doi.org/10.7302/vtaz-tw05, 2022..

Palle, E., Ford, E. B., Seager, S., Montanes-Rodriguez, P., & Vazquez, M., Identifying the Rotation Rate and the Presence of Dynamic Weather on Extrasolar Earth-like Planets from Photometric Observations, *Ap. J.*, 676, 1319-1329, 2008.

Pawlowski, D. and A. Ridley, Modeling the thermospheric response to solar flares, *J. Geophys. Res.* **113**, A10309, 2008.

Pawlowski, D. and A. Ridley, Modeling the ionospheric response to the 28 October 2003 solar flare due to coupling with the Thermosphere, *Radio Sci.* **44**, RS0A23, 2009a.

Pawlowski, D. and A. Ridley, Quantifying the effect of thermospheric parameterization in a global model, *J. Atmos. Sol., Terr. Phys.* **71**, 2017, 2009b.

Pernice et al., Laboratory evidence for a key intermediate in the Venus atmosphere: Peroxychloroformyl radical, *PNAS*, 101, 14007–14010, 1994.

Petrass, J., Honours thesis, Australia National University, Oct 2013.

Pickles, A. J., PASP 110, 863, 1998.

Pidhorodetska, D., et al., *Ap. J. Lett.* **898** L33, 2020.
Prinn, R. G., *J. Atmos. Sci*. 28, 1058–1068, 1971.
Planavsky, N. J., Reinhard, C. T., Wang, X., et al., *Science*, 346, 635, 2014.
Pollack, J.B. et al., Distribution and source of the UV absorption in Venus' atmosphere. *J. Geophys. Res.,* 85, 8141–8150, 1980.
Quintana, E. V., Barclay, T., Raymond, S. N., et al., *Science*, 344, 277, 2014
Ramirez, R. M., & Kaltenegger, L., *Ap. J. Lett*., 797, L25, 2014.
Rauer, H., Catala, C., Aerts, C., Appourchaux, T., Benz, W., Brandeker, A., Christensen-Dalsgaard, J., Deleuil, M., Gizon, L., Goupil, M.-J., Güdel, M., et al. The PLATO 2.0 mission. *Exp. Astron*., 38:249– 330. 2014, DOI: 10.1007/s10686-014-9383-4
Ribas, Ignasi & Font-Ribera, Andreu & Beaulieu, Jean-Philippe. (2008). A ~5 M_earth Super-Earth Orbiting GJ 436?: The Power of Near-Grazing Transits. The Astrophysical Journal Letters. 677. 10.1086/587961. Ricker, G. R., Winn, J. N., Vanderspek, R., Latham, D. W., Bakos, G.Á., Bean, J. L., Berta-Thompson, Z. K., Brown, T. M., Buchhave, L., Butler, N.R., Butler, R. P., et al. Transiting Exoplanet Survey Satellite (TESS). J. Astron. Telescope Instrum. Sys., 1(1):014003., 2015, doi: 10.1117/1.JATIS.1.1.014003
Robinson, T. D., and Catling, D. C., An Analytic Radiative-Convective Model for Planetary Atmospheres," *Ap. J.,* 757, 104, 2012.
Robinson, Tyler D. and Crisp, D., Linearized Flux Evolution (LiFE): A technique for rapidly adapting fluxes from full-physics radiative transfer models, *JQSRT*, 211, 78-95, 2018.
Roldán,C., López-Valverde, M.A., López-Puertas, M.,Edwards, D.P., Non-LTE infrared emissions of CO2 in the atmosphere of Venus. *Icarus,* 147, 11–25, 2000. http://dx.doi.org/10.1006/icar.2000.6432.
Rosenqvist, J., E. Lellouch, T. Encrenaz, and G. Pauber, Global circulation in Venus' mesosphere from IRAM CO observations (1991-1994): A tribute to Jan Rosenqvist, *Bull. Amer. Astro. Soc., 27*, 1080, 1995.
Sandor B. J. and R. T. Clancy, *Icarus*, 290, 156-161, 2017.
Schubert, G., S. W. Bougher, A. D. Covey, C. C. Del Genio, A. S. Grossman, J. L. Hollingsworth, S. S. Limaye, and R. E. Young, Venus atmosphere dynamics: A continuing enigma, in Exploring Venus as Terrestrial Planet, Geophys.Monogr. Ser., vol. 176, edited by L. W. Esposito, E. R. Stofan, and T. E. Cravens, pp. 121–138, AGU, Washington, D. C., 2007.
Schubert, G., et al., Structure and circulation of the Venus atmosphere, *J. Geophys. Res*., 85, 8007–8025, doi:10.1029/JA085iA13p08007, 1980.
Schumann, U. and Huntrieser, H.: The global lightning-induced nitrogen oxides source, *Atmos. Chem. Phys*., 7, 3823-3907, https://doi.org/10.5194/acp-7-3823-2007, 2007.
Seager, S., & Hui, L., Constraining the Rotation Rate of Transiting Extrasolar Planets by an Oblateness Measurement, *Ap. J.*, 574, 1004-1010, 2002.
Seager, S., Bains, W., Hu, R., Biosignature Gases in H2-Dominated Atmospheres on Rocky Exoplanets, *Ap. J.*, 777, 95, 2013.
Segura, A., Kasting, J. F., Meadows, V., et al., *Astrobiology*, 5, 706, 2005.
Segura, A., Meadows, V. S., Kasting, J. F., et al., *A&A*, 472, 665, 2007.
Selsis, F., Despois, D., & Parisot, J.-P., *A&A*, 388, 985, 2002.
Showman, A., & Polvani, L., *GRL*, 37, L18811, 2010.
Showman, A., & Polvani, L., *Ap. J.*, 738, 71, 2011.
Showman, A., Wordsworth, R., Merlis, T., & Kaspi, Y., in Comparative Climatology of Terrestrial Planets, ed. S. Mackwell et al. (Tucson, AZ: Univ. Arizona Press), 277, 2013b.

Sobel, A., Nilsson, J., & Polvani, L., *J. Atmos. Sci*., 58, 3650, 2001.
Simpson, D., et al., *JGR* 104, 8113-8152, 1999.
Smith, M.D., *Icarus* **167**, 148, 2004.
Smith, M.D., *Icarus* **202**, 444, 2009.
Solomon, S., *Reviews of Geophysics,* 37, 275-316, 1999.
Stevenson, K. B., Harrington, J., Lust, N. B., Lewis, N. K.; Montagnier, G., Moses, J. I., Visscher, C., Blecic, J., Hardy, R. A., Cubillos, P., Campo, Christopher J., Two nearby Sub-Earth-sized Exoplanet Candidates in the GJ 436 System, *Ap. J.*, 755, 1, 2012. doi:10.1088/0004-637X/755/1/9.
Stolarski, R. S., and R. J. Cicerone, Stratospheric Chlorine: a Possible Sink for Ozone, *Canadian Journal of Chemistry*, 1974, 52(8): 1610-1615, https://doi.org/10.1139/v74-233
Sze, N. D. & McElroy, M. B., *Planet. Space Sci*., 23, 763–786, 1975.
Tian, F., & Ida, S., *Nature Geo*., 8, 177, 2015.
Tian, F., *E&PSL*, 432, 126, 2015.
Tian, F., France, K., Linsky, J. L., Mauas, P. J., & Vieytes, M. C., *E&PSL*, 385, 22, 2014.
Tsai, S., Dobbs-Dixon, I., & Gu, P., *Ap. J*., 793, 141, 2014.
Turbet M., Bolmont E., Leconte J., et al., *A&A* 612 A86, 2018.
Turbet, M., Bolmont, E., Chaverot, G. *et al.* Day–night cloud asymmetry prevents early oceans on Venus but not on Earth. *Nature,* **598,** 276–280, 2021), https://doi.org/10.1038/s41586-021-03873-w.
Vandaele, A.C., O. Korablev, D. Belyaev, S. Chamberlain, D. Evdokimova, Th. Encrenaz, L. Esposito, K.L. Jessup, F. Lefèvre, S. Limaye, A. Mahieux, E. Marcq, F.P. Mills, F. Montmessin, C.D. Parkinson, S. Robert, T. Roman, B. Sandor, A. Stolzenbach, C. Wilson, V. Wilquet, Sulfur dioxide in the Venus atmosphere: I. Vertical distribution and variability, *Icarus*, Volume 295, October 2017, Pages 16-33, ISSN 0019-1035, https://doi.org/10.1016/j.icarus.2017.05.003. (http://www.sciencedirect.com/science/article/pii/S0019103516303281), 2017 a.
Vandaele, A.C., O. Korablev, D. Belyaev, S. Chamberlain, D. Evdokimova, Th. Encrenaz, L. Esposito, K.L. Jessup, F. Lefèvre, S. Limaye, A. Mahieux, E. Marcq, F.P. Mills, F. Montmessin, C.D. Parkinson, S. Robert, T. Roman, B. Sandor, A. Stolzenbach, C. Wilson, V. Wilquet, Sulfur dioxide in the Venus Atmosphere: II. Spatial and temporal variability, *Icarus*, Volume 295, October 2017, Pages 1-15, ISSN 0019-1035, https://doi.org/10.1016/j.icarus.2017.05.001. (http://www.sciencedirect.com/science/article/pii/S0019103516303293), 2017b.
Valeille, A., M.R. Combi, V. Tenishev, S.W. Bougher, A.F. Nagy. A study of suprathermal oxygen atoms in Mars upper thermosphere and exosphere over the range of limiting conditions, *Icarus*, 206, pp. 18-27, 2010, 10.1016/j.icarus.2008.08.018
Valeille, A., M.R. Combi, V. Tenishev, S.W. Bougher, A.F. Nagy. Three-dimensional study of Mars upper thermosphere/ionosphere and hot oxygen corona: *J. Geophys. Res*., 114, E11006, 2010, doi:10.1029/2009JE003389.
Wallace, M. W., Hood, A. V. S., Shuster, A., et al., *E&PSL*, 466, 12, 2017.
Way, M.J., et al. (2016). *Geophys. Res. Lett*., 43, 8376, 2016.
Way, M.J., Del Genio, A.D. (2020). *JGR: Planets*, 125, 06276, 2020.
West, A. A., Hawley, S. L., Walkowicz, L. M., et al., *Ap. J*., 128, 426 WMO (World Meteorological Organization) 1985, Atmospheric Ozone Assessment of our Understanding of the Process Controlling its Present, Distribution and Change, Vol. 3, WMO Report No.

16 (Geneva: WMO), 2004.

Willacy, K., M. Allen, and Y. Yung, A new astrobiological model of the atmosphere of Titan. *Ap. J.*, 829, 79, 2016.

Wofsy, S. C., and M. B. McElroy, $HO_x$, $NO_x$, and $ClO_x$: Their Role in Atmospheric Photochemistry *Canadian Journal of Chemistry*, 52(8): 1582-1591, 1974, https://doi.org/10.1139/v74-230

Wordsworth, R. D., Forget, F., Selsis, F., et al., *Ap. J. Lett.*, 733, L48, 2011.

Wordsworth, R., & Pierrehumbert, R., *Ap. J. Lett.*, 785, L20, 2014.

Yung, Y. L., and W. B. Demore, Photochemistry of the stratosphere of Venus – Implications for atmospheric evolution, *Icarus*, *51*, 199-247, 1982.

Yung, Y. L., & DeMore, W. B., *Photochemistry of Planetary Atmospheres* (New York, NY: Oxford Univ. Press), 1999.

Zahnle, K., Haberle, R. M., Catling, D. C., et al., *JGR*, 113, E11004, 2008.

Zurek, R. W., et al., *JGR: Space Physics*, 122, doi:10.1002/2016JA023641, 2017.

Zhang, X., M.-C. Liang, F. Montmessin, J.-L. Bertaux, C. Parkinson, and Y.L. Yung. Photolysis of sulfuric acid as the source of sulfur oxides in the mesosphere of Venus. *Nature Geoscience*, 3. 834–837. DOI: 10.1038/NGEO989, 2010.

Zhang, S., M.-C. Liang, F. Mills, D. A. Balyaev, and Y. L. Yung, Sulfur Chemistry in the Middle Atmosphere of Venus, *Icarus*, 217, 714-739, 2012.